\documentclass[preprint,12pt,authoryear]{elsarticle}

\usepackage{amssymb}
\usepackage{amsmath} 
\usepackage{graphicx}
\usepackage{subfig}
\usepackage{wasysym}
\usepackage{ltxtable}
\usepackage{tabularx}
\usepackage{longtable}
\usepackage{lipsum}
\usepackage{lscape}
\usepackage{color}
\usepackage{epsf}
\usepackage{hyperref}
\usepackage{aas_macros}
\usepackage{ulem}
\usepackage{bm}
\usepackage{geometry}
\newcounter{RomanNumber}

\newcommand{\tabincell}[2]{\begin{tabular}{@{}#1@{}}#2\end{tabular}}
\journal{Journal of High Energy Astrophysics}

\begin{document}

\begin{frontmatter}

%% Title, authors and addresses

%% use the tnoteref command within \title for footnotes;
%% use the tnotetext command for theassociated footnote;
%% use the fnref command within \author or \address for footnotes;
%% use the fntext command for theassociated footnote;
%% use the corref command within \author for corresponding author footnotes;
%% use the cortext command for theassociated footnote;
%% use the ead command for the email address,
%% and the form \ead[url] for the home page:
%% \title{Title\tnoteref{label1}}
%% \tnotetext[label1]{}
%% \author{Name\corref{cor1}\fnref{label2}}
%% \ead{email address}
%% \ead[url]{home page}
%% \fntext[label2]{}
%% \cortext[cor1]{}
%% \address{Address\fnref{label3}}
%% \fntext[label3]{}

\title{Possible correlations between gamma-ray burst and its host galaxy offset}

%% use optional labels to link authors explicitly to addresses:
%% \author[label1,label2]{}
%% \address[label1]{}
%% \address[label2]{}

\author[hust]{Fei-Fei Wang}

\author[hust]{Yuan-Chuan Zou\corref{cor}}
\ead{zouyc@hust.edu.cn}

\author[hust]{Yu Liu}

\author[hust]{Bin Liao}

\author[racah]{Reetanjali Moharana}

\cortext[cor]{Corresponding author}

\address[hust]{School of Physics, Huazhong University of Science and Technology,  Wuhan 430074, China}

\address[racah]{Racah Institute of Physics, The Hebrew University, Jerusalem 91904, Israel}

\begin{abstract}
We collected the information of 304 gamma-ray bursts (GRBs) from the literature, and analyzed the correlations among the host galaxy offsets (the distance from the site of the GRB to the center of its host galaxy), $T_{\rm 90,i}$ (the duration $T_{\rm 90}$ in rest-frame), $T_{\rm R45,i}$ (the duration $T_{\rm R45}$ in rest-frame), $E_{\rm \gamma,iso}$ (the isotropic equivalent energy), $L_{\rm \gamma,iso}$ ($=E_{\rm \gamma,iso}/T_{\rm 90,i}$, the isotropic equivalent luminosity) and $L_{\rm pk}$ (peak luminosity). We found that $ T_{\rm 90,i}$, $ T_{\rm R45,i}$, $ E_{\rm \gamma,iso}$,  $L_{\rm pk}$ have negative correlation with $ {\rm offset}$, which is consistent with origin of short GRBs (SGRBs) and long GRBs (LGRBs). On separate analysis, we found similar results for $\log E_{\rm \gamma,iso}$ - $\log {\rm (offset)}$ and $\log L_{\rm pk}$ - $\log {\rm (offset)}$ relations in case of SGRBs only, while no obvious relation for LGRBs. There is no correlations between offset and $L_{\rm \gamma,iso}$. We also put the special GRB 170817A {and GRB 060218A} on the plots. {The two GRBs both have low luminosity and small offset.} In the  $ \log ({\rm offset})- \log T_{\rm 90,i}$ plot, we found  GRB 170817A locates in between the two regions of SGRBs and LGRBs  and it is the outlier in the $ {\rm offset}-E_{\rm \gamma, iso}$, $ {\rm offset}-L_{\rm \gamma, iso}$ and $ {\rm offset}-L_{\rm pk}$ plots.  Together with GRB 060218A being an outlier in all plots, it indicates the speciality of GRBs 170817A and 060218A, and might imply more subgroups of the GRB samples.
\end{abstract}

\begin{keyword}
gamma-ray bursts \sep statistics \sep host galaxy offsets \sep $T_{\rm 90}$ \sep $T_{\rm R45}$ \sep $E_{\rm \gamma,iso}$
%% keywords here, in the form: keyword \sep keyword

%% PACS codes here, in the form: \PACS code \sep code
\PACS 98.70.Rz \sep 98.52.-b
%% MSC codes here, in the form: \MSC code \sep code
%% or \MSC[2008] code \sep code (2000 is the default)

\end{keyword}

\end{frontmatter}

%% \linenumbers

%% main text
\section{Introduction}
\label{sec:intro}

%% The Appendices part is started with the command \appendix;
%% appendix sections are then done as normal sections
%% \appendix

%% \section{}
%% \label{}

%% If you have bibdatabase file and want bibtex to generate the
%% bibitems, please use
%%
%%  \bibliographystyle{elsarticle-harv} 
%%  \bibliography{<your bibdatabase>}

%% else use the following coding to input the bibitems directly in the
%% TeX file.
Gamma-ray bursts (GRBs) are widely accepted to have two categories, short GRBs (SGRBs) having duration shorter than 2 $\rm s$ and long GRBs (LGRBs) with duration longer than 2 $\rm s$ \citep{Kouveliotou1993}. SGRBs are thought to be from the merger of compact object binaries involving at least one neutron star \citep{Eichler1989,Paczynski1991,Narayan1992}, and have a broad range of spatial host galaxy distribution \citep{Zhang2017}. The origin of LGRBs are most-likely to be the collapse of rapidly-rotating massive stars \citep{MacFadyen1999}, hence expected to be inside the star forming region. Consequently, the offsets of the location in the host galaxy of LGRBs are mostly smaller than those of SGRBs.

In the past few decades, there have been many studies on host galaxy offsets of GRBs. For example, \citet{Bloom2002} studied host galaxy offsets for LGRBs. The result was consistent with the expected distribution of massive stars, confirming the core-collapse model as the origin of LGRBs.  \citet{Fong2010A} presented the first comprehensive analysis of $Hubble ~ Space ~ Telescope ~ (HST)$ observations of ten SGRBs host galaxies. Their result showed an median at 5 $\rm kpc$ for SGRBs host galaxy offsets, which is about 5 times larger than LGRBs. There was no evidence of differences between SGRBs with and without extended emission. The host galaxy offsets are in good agreement with neutron star binary mergers (see also \citet{Church2011}). However, \citet{Malesani2007} noticed that SHBs (short hard GRBs) with extended emission are more easier to detect their optical counterparts. This has been explained as an environmental property by \citet{Troja2008}, as SHBs with extended emission seem to occur closer to their host galaxies, in denser interstellar environments. This also implies that SGRBs progenitors have an intrinsically different behavior, due to their association with different origins such as black hole (BH)-neutron star (NS) and NS-NS merger. \citet{Troja2008} showed that SGRBs with extended hard X-ray emissions that have small projected physical offsets may be due to NS-BH mergers, while those without extended hard X-ray emission components that have bigger projected physical offsets may be due to NS-NS mergers. The correlation between X-ray absorption column densities and host galaxy offsets gives another evidence that SGRBs possibly have two distinct populations \citep{Kopac2012}. Furthermore, some negative correlations are found between the broadband afterglow emissions and SGRBs host galaxy offsets \citep{Zhang2017}. This is because the afterglow emission depends on the circum-burst medium and it decreases with the distance to the host galaxy center, providing more evidences that SGRBs with larger host galaxy offsets prefer lower circum-burst densities \citep{Fong2015}.

To investigate the properties of the host galaxies and the connection to the GRBs, we collect all the possible sample from the literature about the offsets, durations of the GRBs ($T_{90}$ (time duration from 5\% photon counts to 95\% photon counts) and $T_{\rm R45}$  \citep[defined in][]{Reichart2001}),  the isotropic equivalent $\gamma-$ray energy $E_{\rm \gamma, iso}$, and the 1 s time binned peak luminosity $L_{\rm pk}$. In this work we analyze these data and present our results for the relations found for SGRBs, LGRBs and combination of them.  The paper is organized as the follow: the data is collected and described in \S \ref{sec:samples}, the statistics is performed in \S \ref{sec:results}, and conclusion and discussion is given in \S \ref{sec:discussion}.

\section{The GRB sample}
\label{sec:samples}

We selected 304 GRBs from different instruments, and collected their trigger time, instrument, redshift $z$, offset, $T_{\rm 90}$, $T_{\rm R45}$, $E_{\rm \gamma,iso}$ and $L_{\rm pk}$ values from different published papers. All the information is provided in Table \ref{tab:sample} to \ref{tab:lpklast}. $E_{\rm \gamma,iso}$ and $L_{\rm pk}$ are in rest-frame 1-$10^{4}$ $\rm keV$ energy band, and $L_{\rm pk}$ is in 1 $\rm s$ time bin (except GRB 170817A in 50 $\rm ms$ time bin). We also calculated isotropic equivalent luminosity $L_{\rm \gamma,iso}$ in the rest-frame 1-$10^{4}$ $\rm keV$ energy band, which is $ L_{\rm \gamma,iso}=(1+z)E_{\rm \gamma,iso}/T_{90}$. For $L_{\rm pk}$, sometimes the energy band is not in rest-frame 1-$10^{4}$ $\rm keV$ energy band, like \citet{Deng2016}. We changed the energy band using the spectral information. There are mainly three kinds of spectral models: Band model, cutoff power law (CPL) model and simple power law (SPL) model \citep[more details in][]{Li2016ApJS}.  In Table \ref{tab:lpk} to \ref{tab:lpklast}, we gave the GRB spectral information which need to change the energy band, as well as the $L_{\rm pk}$. For Band model, $\alpha$, $\beta$ and $E_{\rm pk}$ are low energy spectral index, high energy spectral index and peak energy, respectively. For CPL model, $\alpha$ is the spectral index for the power law band and $E_{\rm pk}$ is the cutoff energy. There is no $\beta$ for CPL model, and we use ``..." to remark $\beta$. %The formula of SPL model is $N(E) = A E^{\alpha}$, $\alpha$ is the spectrum index of SPL model. We use ``..." to remark $\beta$ and $E_{\rm pk}$ of SPL model. 
Besides, we excluded some values with lower limit smaller than 0. For example, the offset of GRB 120119A is $0.104 \pm 0.147$ \citep{Li2016ApJS}.

The data are not complete, as not every GRB has all the observational values listed above, available. Some of the data have only the central values available without error bars. To keep the information of the central values, we need to impute the errors from other data. We used the R package $mice$ to impute the error bars for the data that have just the central values, by multiple imputation with chained equations (MICE) \citep{Rubin1987,Rubin1996}. 

\subsection{Error imputation} \label{subsec:imputation}
We use the R package $mice$ to impute incomplete multivariate data by using the method MICE. MICE is a powerful tool for imputation and it has been widely used. Only the central values with missing error bars are imputed. The ones with missing central values are omitted in the statistical analysis. According to \citet{Rubin1987,Rubin1996}, MICE includes three steps: generating multiple imputation, analyzing imputed data, and pooling analysis results. 

The imputation model should also have three principles: accounting for the process that created the missing data, preserving the relations in the data and preserving the uncertainty about these relations.  At first, we changed $T_{\rm 90,i}$, $T_{\rm R45,i}$, $E_{\rm \gamma,iso}$ and host galaxy offset into their logarithmic values. Then we did 5 times imputation as suggested in \citet{Rubin1996}. It means every error bar which need to be imputed will have 5 imputed values, hence we have 5 complete set of data. We need to choose the imputation model first, because our data is missing at random (MAR) \citep{Rubin1976}, additionally, our data is numeric type. So we choose the predictive mean matching model (PMM) \citep{Little1988}, a general purpose semi-parametric imputation method \footnote{We compared the correlations between the central values and the related errors, and found the PMM is reliable in the error imputation. For example, the positive error of $T_{\rm 90,i}$ is $T_{\rm 90,i,1}$. Before imputation, the linear regression between $T_{\rm 90,i}$ and $T_{\rm 90,i,1}$ is $T_{\rm 90,i} = (1.26 \pm 0.05) + (-9.28 \pm 1.58) \times T_{\rm 90,i,1}$, and the Pearson coefficient is $-0.33 \pm 0.05$ with p-value $1.2 \times 10^{-8}$. After the imputation, the linear regression between $T_{\rm 90,i}$ and $T_{\rm 90,i,1}$ is $T_{\rm 90,i} = (1.27 \pm 0.05) + (-9.7 \pm 1.58) \times T_{\rm 90,i,1}$, and the Pearson coefficient is $-0.33 \pm 0.05$ with p-value $1.6 \times 10^{-9}$. The results do not change too much, which means PMM model can preserve the relations in the data and preserve the uncertainty about these relations.}. We set a threshold at 0.25, which means the minimum proportion of usable cases for imputation is at least 0.25. An important step in multiple imputation is that, we want to assess whether imputations are plausible, then we have done diagnostic checks. We used following three indicators to assess the goodness of our imputation results.

\begin{enumerate}
\item Relative increase in variance due to missing data $r_{\rm m}$ (RIV): It is the ratio between imputation variance and the imputation variance of the 5 data sets, then multiplying the imputation time m. It stands for the increase fraction in variance due to missing data, the influence of the missing data is bigger when $r_{\rm m}$ is bigger. While smaller $r_{\rm m}$ indicates influence of the change of m is smaller, this is to say that missing data has smaller influence to the whole data parameters, hence the imputation results are more stable and the imputations are better.

\begin{equation} \label{eq:RIV}
r_{m}=\frac{(1+\frac{1}{m}) {\sigma_{\rm B}}^{2}}{{\sigma_{\rm W}}^{2}}.
\end{equation}

${\sigma_{\rm W}}^{2}$ is within-imputation variance, it represents the mean of the variance for the m data sets.

\begin{displaymath}
{\sigma_{\rm W}}^{2}=\frac{1}{m} \sum_{i=1}^m {\sigma_{\rm i}}^{2}.
\end{displaymath}

${\sigma_{\rm B}}^{2}$ is between-imputation variance, it represents the variance of the mean of m data sets.

\begin{displaymath}
{\sigma_{\rm B}}^{2}=\frac{1}{m-1} \sum_{i=1}^m {(\widehat{\theta}_{\rm i}-\widehat{\theta})}^{2}
\end{displaymath}

$\widehat{\theta}_{\rm i}$ is the mean of every complete data set, $\widehat{\theta}=\frac{1}{m}\sum_{i=1}^m \widehat{\theta}_{\rm i}$

\item Fraction of missing information $\gamma_{\rm m}$ (FMI): This represents the influence of the missing data for the whole parameters(e.g. mean). Smaller FMI values indicate that the imputation results are more stable.

\begin{equation} \label{eq:FMI}
\gamma_{\rm m}=\frac{r_{\rm m}+\frac{2}{v_{\rm m}+3}}{r_{\rm m}+1}
\end{equation}

$v_{\rm m}=(m-1)(1+\frac{1}{r_{\rm m}}^{2})$ is the degree of freedom. 

\item Relative efficiency (RE): is a comprehensive analysis of RIV and FMI. It represents the imputation fraction for missing information by MICE. The higher value of RE means the better result. 

\begin{equation} \label{eq:RE}
{\rm RE}={(1+\frac{\gamma_{\rm m}}{m})}^{-1}
\end{equation}

\end{enumerate}

For analyzing imputed data and pooling analysis results, we use the mean of every imputed error bar, because we also need to calculate some values and plot scatter plots with error bars.  As there are 5 candidate values for each parameter, we use the mean of them as the imputed error.

The imputation results are shown in  Table \ref{tab:impu}. From the results, we can see that RIV and FMI are very close to 0, which means our imputation is stable. RE is very close to 1, which means our imputation efficiency is very high, as per the definition. We almost imputed all the missing information. Therefore, we justify the imputation is reliable.

\section{Correlation method and results} \label{sec:results}
Due to the special features of GRB 170817A and GRB 060218A, we have not included them in our statistical analysis. To figure out the intrinsic connections between different properties associated with the GRBs, we convert the values into rest frame. We changed $T_{\rm 90}$ and $T_{\rm R45}$ from observer-frame to rest-frame by dividing them with $1+z$ and assigned them as $T_{\rm 90,i}$ and $T_{\rm R45,i}$ respectively. At first, we analyzed using the combined sample of both SGRBs and LGRBs. Then we separated the GRBs into SGRB group and LGRBs group, and did the same statistical analysis for the two groups separately. 

In order to make sure the results are reliable, we have used several statistical analysis narrated below. At first, we calculated the correlation coefficients to check the linear correlations between different parameters, with Pearson, Spearman and Kendall $\tau$ correlation methods \citep{Feigelson2012}. Pearson correlation coefficient is to measure whether the two parameters are aligned in the two-dimension plot. Spearman and Kendall $\tau$ correlation coefficients are methods to measure the monotony of the two parameters. Then we did the linear regression to get the parameters of the linear expression. 

For all the correlation coefficients and linear regression, we have done hypothetical tests. Once the results pass through the hypothetical tests with p-value smaller than 0.1 we accepted the correlation, as this value gives a high probability of relying the result. In the correlation statistics we have considered the error following the method used in \citet{Zou2017}. Assuming the errors are normally distributed, we generated  $10^{\rm 3}$ sets of random samples with Monte Carlo (MC) simulation. Then we have done the above mentioned statistics for the $10^{\rm 3}$ sets and checked their distribution. The best fit and corresponding 1$\sigma$ errors are finally obtained.

Using above mentioned methods we found significant correlations for the following pairs, $\log {\rm (offset)} - \log T_{\rm 90,i}$,  $\log {\rm (offset)}  - \log T_{\rm R45,i}$,  $\log {\rm (offset)}  - \log E_{\rm \gamma,iso}$, $\log {\rm (offset)}  - \log L_{\rm pk}$, and  $\log E_{\rm \gamma,iso}  - \log T_{\rm R45,i}$. The correlation of $\log T_{\rm 90,i}$ and $\log T_{\rm R45,i}$ is also tight. But it is trivial, and we will not discuss about it further. The $\log E_{\rm \gamma,iso}  - \log T_{\rm 90,i}$ relation is not shown here as it has been extensively investigated.
All the correlation coefficient results are listed in Table \ref{tab:coef}. For the combined GRBs, the linear coefficients are significant for all the five relation pairs, the significance of correlation has further been verified with the hypothetical tests for the linear regression coefficients.

\begin{figure*}[!b]
\includegraphics[width=0.4\textwidth]{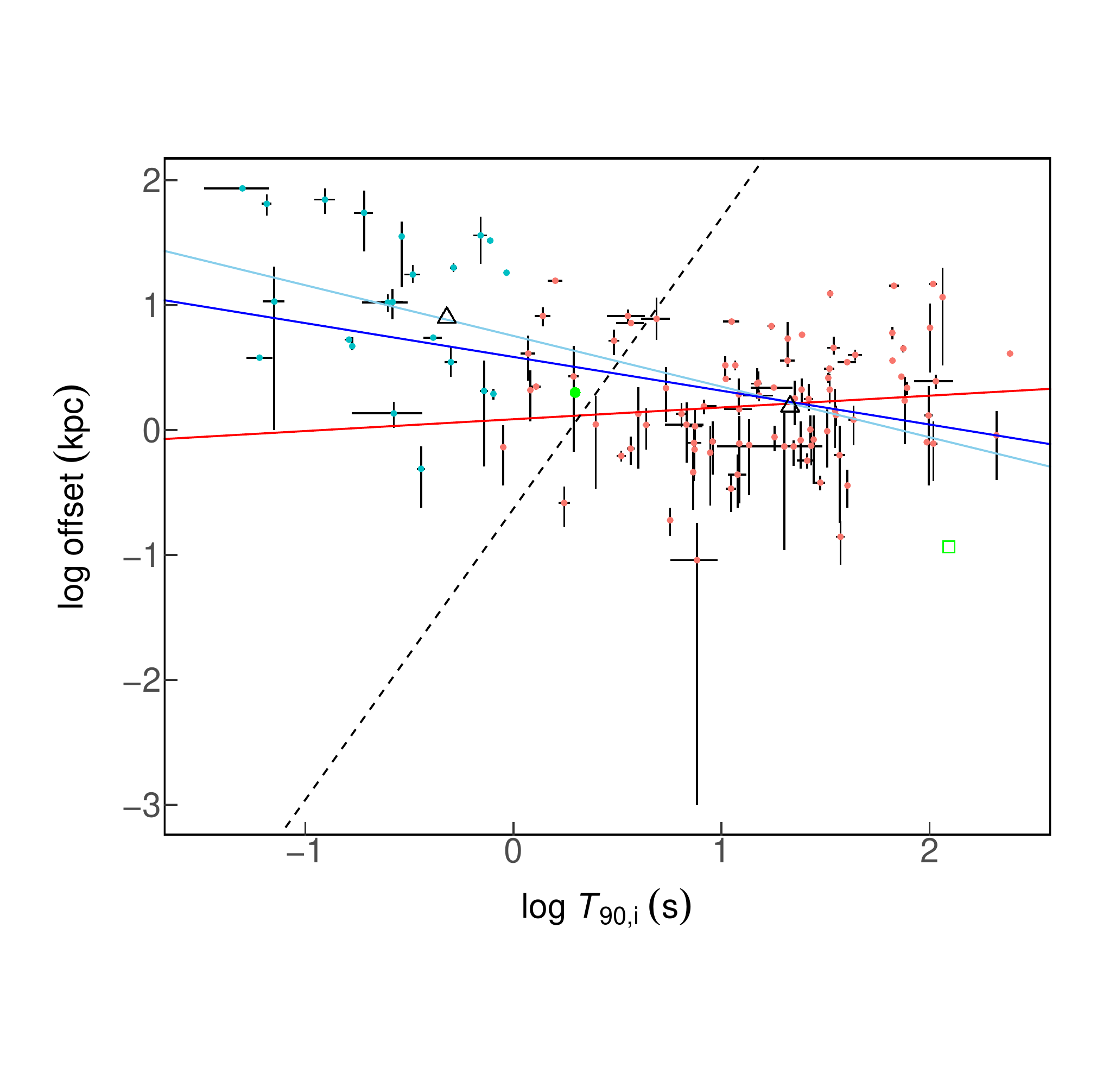}
\includegraphics[width=0.4\textwidth]{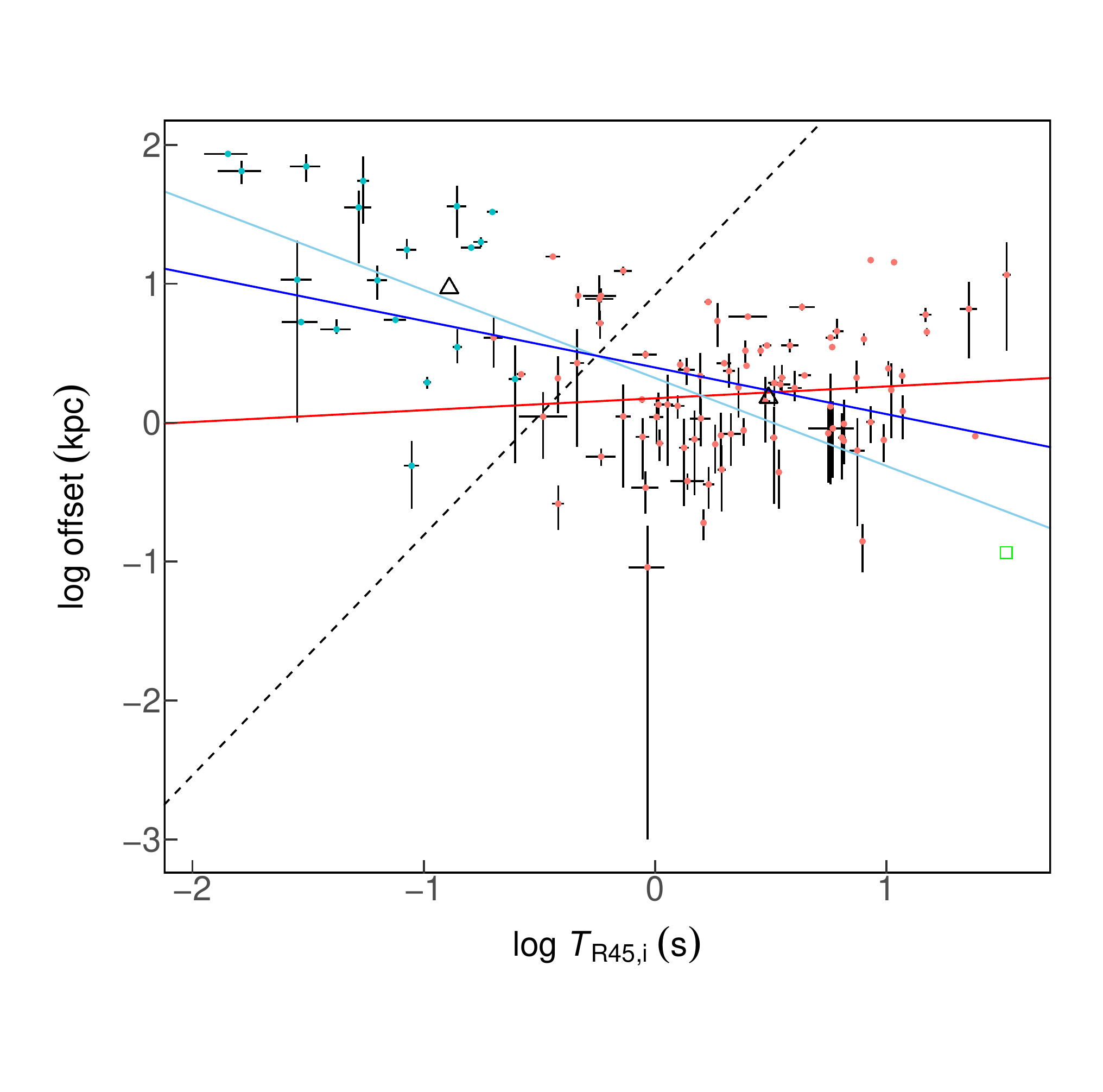} \\
\includegraphics[width=0.4\textwidth]{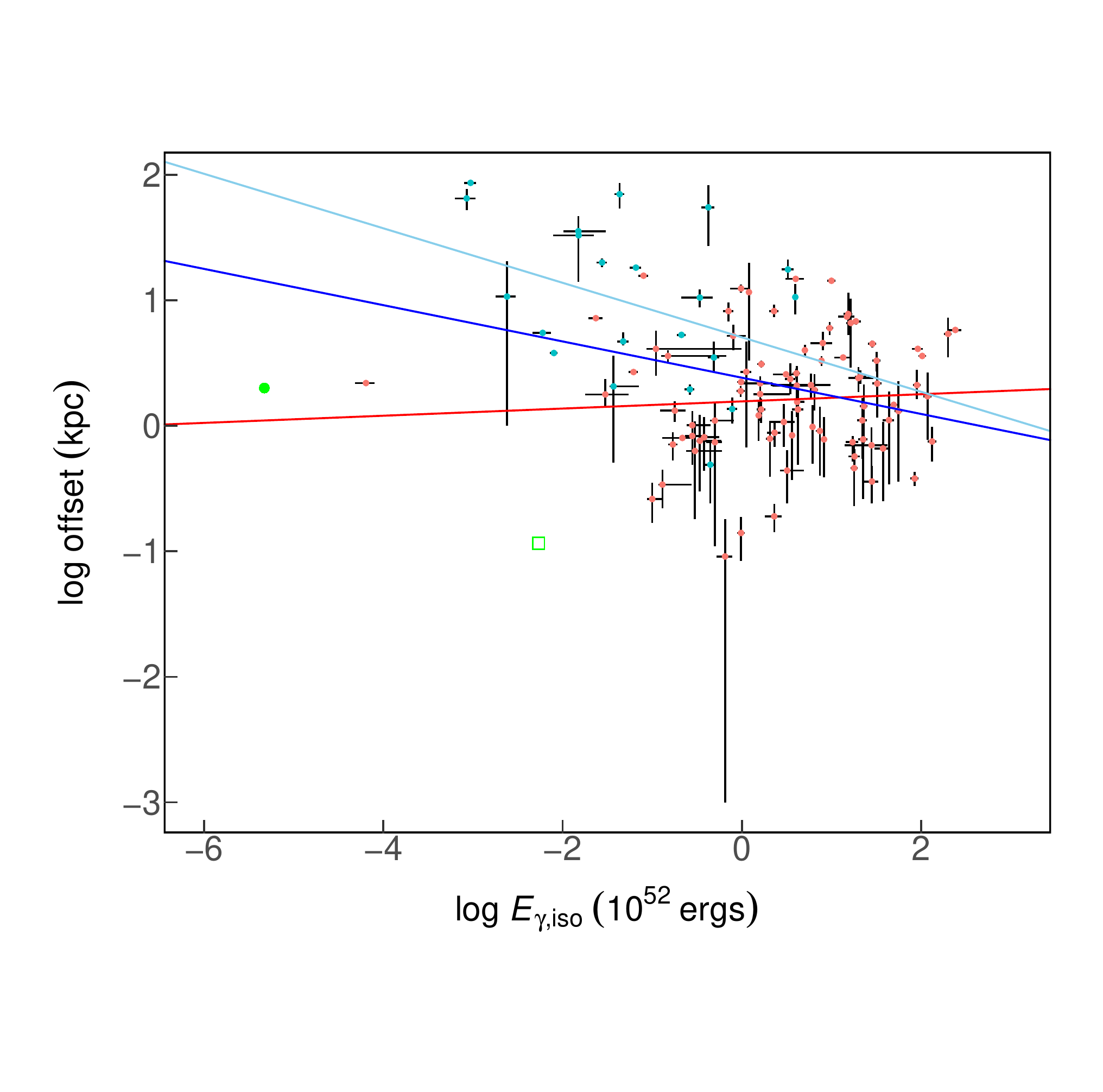}
\includegraphics[width=0.4\textwidth]{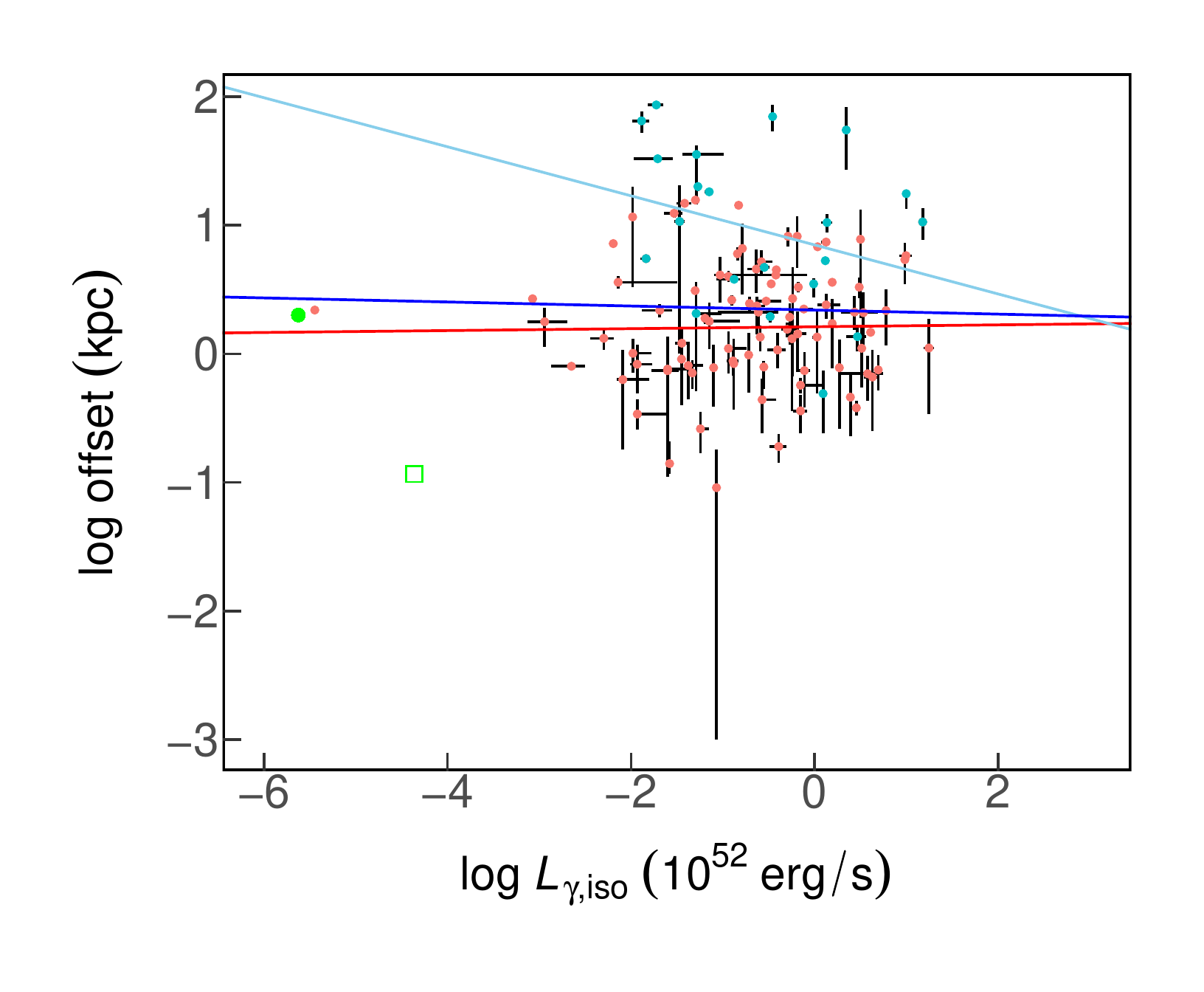} \\
\includegraphics[width=0.4\textwidth]{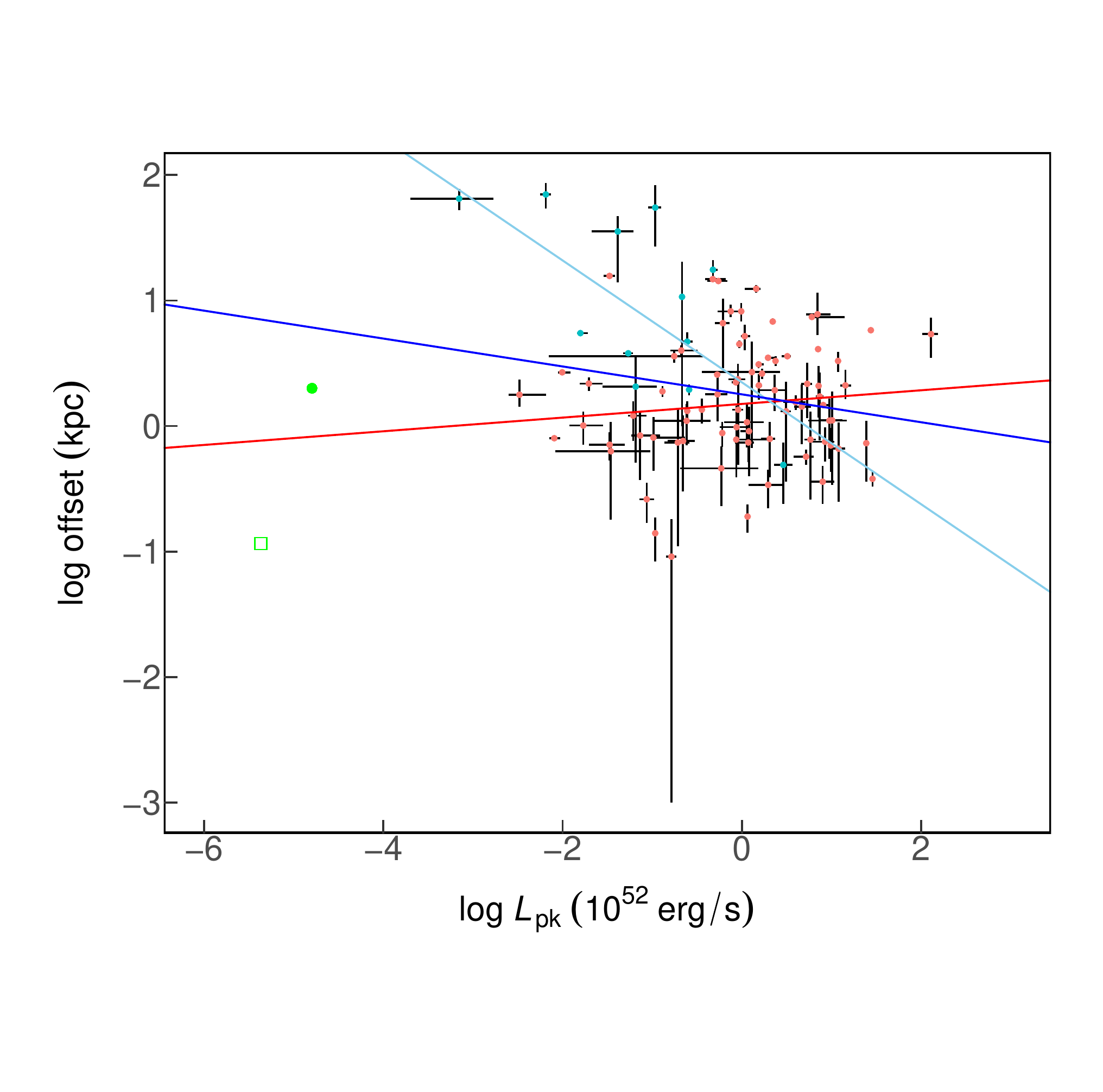}
\includegraphics[width=0.4\textwidth]{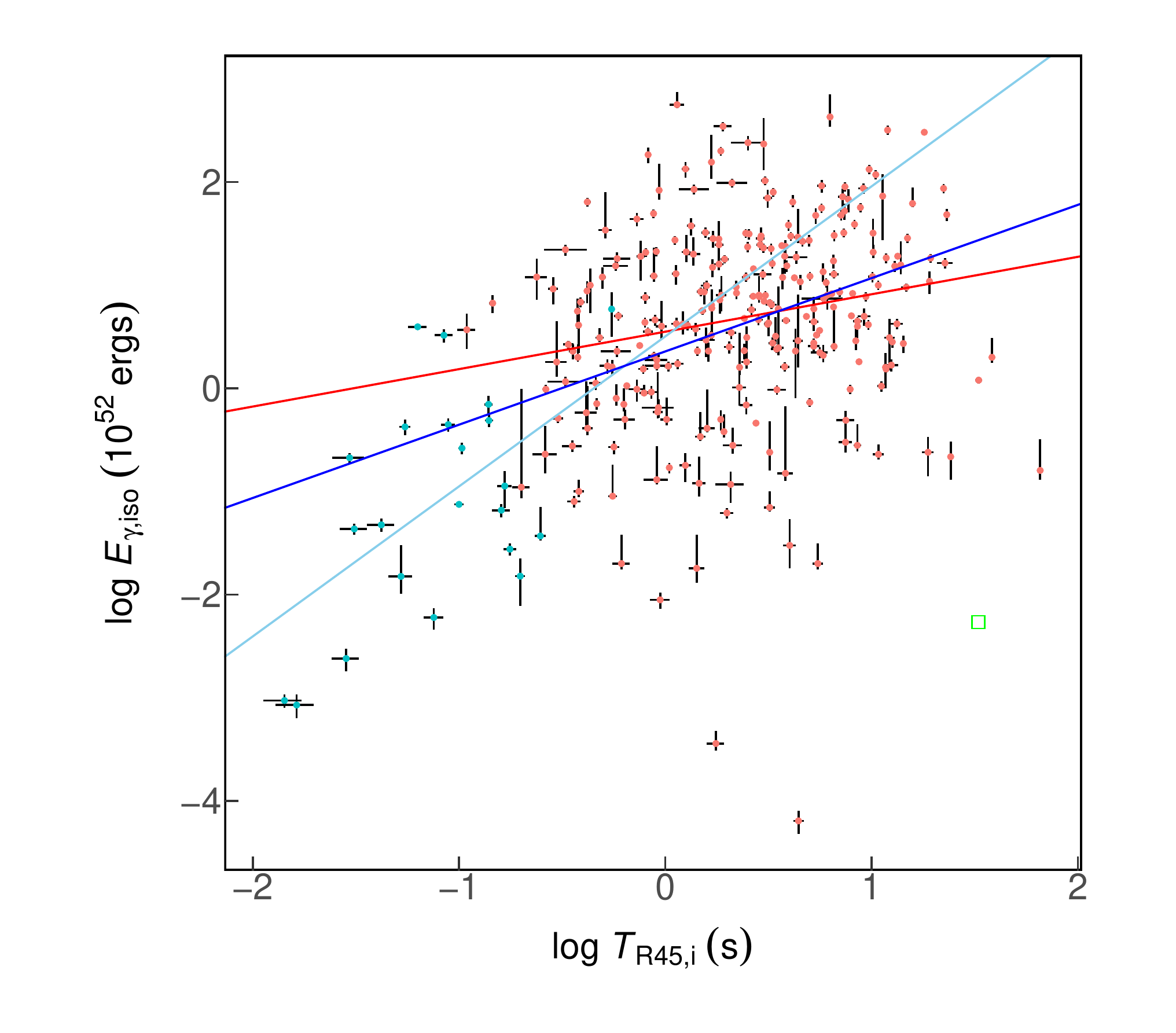}
\caption{The scatter plots and linear regression results between two different parameters, ${\rm offset} - T_{90, \rm i}$,  ${\rm offset} - T_{R45, \rm i}$,  ${\rm offset} - E_{\rm \gamma,iso}$, ${\rm offset} - L_{\rm \gamma,iso}$, ${\rm offset} - L_{\rm pk}$ and  $ E_{\rm \gamma,iso} - T_{\rm R45,i}$ respectively, which are all shown in logarithmic scale. The red points are for LGRBs, and the sky blue points are for SGRBs. The red line is the linear regression result for LGRBs. The sky blue line is the linear regression result for SGRBs. The blue line is the linear regression result for all GRBs. The black dashed line is the result for k-means classification. The two black triangles are the cluster centers. The green circle point is GRB 170817A in ${\rm offset} - T_{90, \rm i}$, ${\rm offset} - L_{\rm \gamma,iso}$, ${\rm offset} - L_{\rm pk}$ and ${\rm offset} - E_{\rm \gamma,iso}$ plots. The green square point is GRB 060218A in all plots. All the linear regression results are derived with MC method similar as in \citet{Zou2017}.}
\label{fig:all}
\end{figure*}

We have shown the scatter plots in Fig. \ref{fig:all}. One can confirm the five correlations intuitively with the scatter plots. In the figures, the fitting line is in blue which was obtained by considering all the errors with MC method similar as in \citet{Zou2017}. The relations obtained are as following,

\begin{equation} \label{eq:allt90offset}
\log {\rm (offset)} = -0.27_{\rm -0.02}^{\rm +0.02} \times \log T_{\rm 90,i} + 0.59_{\rm -0.03}^{\rm +0.01},
\end{equation}

\begin{equation} \label{eq:alltr45offset}
\log {\rm (offset)} = -0.34_{\rm -0.03}^{\rm +0.01} \times \log T_{\rm R45,i} + 0.4_{\rm -0.02}^{\rm +0.01},
\end{equation}

\begin{equation} \label{eq:alleisooffset}
\log {(\rm offset)} = -0.14_{\rm -0.02}^{\rm +0.01} \times \log E_{\rm \gamma,iso,52} + 0.38_{\rm -0.02}^{\rm +0.02},
\end{equation}

\begin{equation} \label{eq:alllpkoffset}
\log {(\rm offset)} = -0.11_{\rm -0.02}^{\rm +0.01} \times \log L_{\rm pk,52} + 0.25_{\rm -0.03}^{\rm +0.03},
\end{equation}

\begin{equation} \label{eq:alltr45eiso}
\log E_{\rm \gamma,iso,52} = 0.71_{\rm -0.01}^{\rm +0.01} \times \log T_{\rm R45,i} + 0.36_{\rm -0.01}^{\rm +0.01},
\end{equation}
where the offset is in unit of kpc, $T_{\rm 90,i}$ and $T_{\rm R45,i}$ are in units of seconds, $L_{\rm pk,52}$ is in unit of $10^{52}$ ergs/s and $E_{\rm \gamma,iso,52}$ is in unit of $10^{52}$ ergs. As expected the $T_{\rm 90,i}$ and $T_{\rm R45,i}$ have similar correlation with the host galaxy offset with slope around $-0.25$. However there is no obvious separation between SGRBs and LGRBs in $T_{\rm 90,i}$, $T_{\rm R45,i}$ and spatial offsets from plots as seen in \citet{Troja2008} for $T_{\rm 90}$ and spatial offset scatter plot. We assume this discrepancy is due to the redshift correction. Interestingly, in the  $\log {(\rm offset)}  - \log T_{\rm 90,i}$ plot, there are four LGRBs with $T_{\rm 90,i}< 2$, and there is a gap between these short ones and the normal LGRBs. This might indicate that the separation of LGRBs and SGRBs should be in the rest frame rather than the observer's frame. The anti-correlation between offset and the duration is consistent with the double degenerated star merger origin for SGRBs and massive star core-collapsing origin for LGRBs. For mergers, they have to pass supernova explosion, which kicks the central compact star. And it takes time for the double compact star to merge, which lead the site of the merging to be in the outer part of the host galaxy. While for the massive star core-collapsing, it should be in the star forming region, which is near the center of the host galaxy. 

The anti-correlation between offset and $E_{\rm \gamma,iso}$, $L_{\rm pk}$ respectively may also be related to the LGRBs and SGRBs, as LGRBs are relatively stronger in $\gamma-$rays, which can also be seen in the  $\log E_{\rm \gamma,iso}  - \log T_{\rm R45,i}$ scattering plot. However we could not find an exact explanation for the values of the slopes and intercepts. We assume a detail analysis of the star forming and merger profile in the galaxies would be able to explain these parameters, however this analysis is beyond the scope of our paper. 

Fig. \ref{fig:all} also gives a clear indication that all the GRBs are gathered in two distinct groups. We apply K-Means clustering algorithms to get the boundary between this two traditionally classes. The boundary is represented with a dash line in  Fig. \ref{fig:all}.  Two different clusters cannot have any object in common. The similarity measure between the cases defined as the euclidean distance on the duration-offset plane. Variables with incompatible units are faced in our dataset. We use logarithmic variables, rather than normalized or standardized variables. Because units can be removed by taking logarithms and the ranges of variables is similar. We kept information on burst duration and offset. The boundaries are $\log({\rm offset} ) = 2.33 \times \log T_{90,i} - 0.63$ and $\log( {\rm offset} ) = 1.73 \times \log T_{\rm R45,i} + 0.92$. We assume that the group of bursts on the duration-offset plane is separated and equal-sized. Each cluster is described by a single point known as the centroid. The centroid of each class is defined as the mean values of $\log  T_{90,i}$, $\log  T_{\rm R45,i}$, and $\log {\rm (offset)}$. The centroids are marked with black open triangles in Fig. \ref{fig:all}. These two groups are generally consistent with long and short GRBs, while with few outliers. 

We have also checked the correlations of the offset with other parameters, while no much significant relations has been found. For the ${\rm offset}-E_{\rm p}$ correlation, as the peak energy can be best fit with Band function or with CPL model, there are two correlations. For the Band function model,  the Pearson coefficient is  $0.30 \pm 0.22$ with p-value 0.32. For the CPL model, the Pearson coefficient is  $0.03 \pm 0.13$ with p-value 0.88. Though there is relatively weak correlation between  ${\rm offset}$ and $L_{\rm pk}$, there is no strong evidence for  ${\rm offset}$ and $L_{\rm \gamma, iso}$. The corresponding Pearson  coefficient is  $-0.03^{+ 0.01}_{-0.03}$ with p-value 0.73 as also shown in Table \ref{tab:coef}.
    
It is important to decide the true number of groups. There is no certain way of telling the goodness of the clustering. The silhouette coefficient is a measure of  the compactness and separation of the clusters. It increases as the quality of the clusters increase, it is large for compact clusters that are far from each other and small for large and overlapping clusters. According to the silhouette coefficient, the most probable number of clusters is 2 for our set of GRBs.

To figure out the correlations are solely caused by the long or short GRBs progenitors, or there is intrinsic tendency inside LGRBs or SGRBs, we divided the 304 GRBs into SGRBs and LGRBs following the widely believed relation of $T_{90}$ with 2s. The coefficients are also listed in the Table \ref{tab:coef}. We list the relations with mean correlation coefficient p-value smaller than 0.1.

The relation between $E_{\rm \gamma,iso}$ and host galaxy offsets for SGRBs is:
\begin{equation} \label{eq:shorteisooffset}
\log ({\rm offset}) = -0.22_{\rm -0.05}^{\rm +0.02} \times \log E_{\rm \gamma,iso,52} + 0.71_{\rm -0.05}^{\rm +0.05},
\end{equation}
where the offset is in unit of kpc, and $E_{\rm \gamma,iso,52}$ is in unit of $10^{52}$ ergs.

The relation between $L_{\rm pk}$ and host galaxy offsets for SGRBs is:
\begin{equation} \label{eq:shortlpkoffset}
\log ({\rm offset}) = -0.49_{\rm -0.08}^{\rm +0.04} \times \log L_{\rm pk,52} + 0.35_{\rm -0.13}^{\rm +0.13},
\end{equation}
where the offset is in unit of kpc, and $L_{\rm pk,52}$ is in unit of $10^{52}$ erg/s.

The relation between $T_{\rm R45,i}$ and $E_{\rm \gamma,iso}$ for SGRBs is:
\begin{equation} \label{eq:shorttr45eiso}
\log E_{\rm \gamma,iso,52} = 1.45_{\rm -0.11}^{\rm +0.05} \times \log T_{\rm R45,i} + 0.5_{\rm -0.12}^{\rm +0.12},
\end{equation}
where $T_{\rm R45,i}$ are in units of seconds, and $E_{\rm \gamma,iso,52}$ is in unit of $10^{52}$ ergs.

The relation between $T_{\rm R45,i}$ and $E_{\rm \gamma,iso}$ for LGRBs is:
\begin{equation} \label{eq:longtr45eiso}
\log E_{\rm \gamma,iso,52} = 0.36_{\rm -0.02}^{\rm +0.01} \times \log T_{\rm R45,i} + 0.55_{\rm -0.01}^{\rm +0.01},
\end{equation}
where $T_{\rm R45,i}$ are in units of seconds, and $E_{\rm \gamma,iso,52}$ is in unit of $10^{52}$ ergs.

When the GRBs are divided into long and short ones, there is no clear correlation between the offset and the duration. It shows that the origin of LGRBs and SGRBs is only due to the spatial offset. The correlations of $\log ({\rm offset})$ - $\log E_{\rm \gamma,iso}$ and $\log({\rm offset} ) - \log  L_{\rm pk}$ observed only for SGRBs. It might indicate that the SGRBs can still be divided into several sub-groups. A possible explanation could be the sub-groups of NS-NS merging and BH-NS merging as origin of SGRBs. For the BH-NS merger, the total gravitational energy is larger than the NS-NS merger, as the BH mass can be much larger. On the other hand, the more massive BH is harder to be kicked outside the host galaxy, and should be located near to the center. This naturally explains the anti-correlation shown in Eq. (\ref{eq:shorteisooffset}). 
Interestingly, as shown in Table \ref{tab:coef} for short GRBs, also for the ${\rm offset}-L_{\rm \gamma, iso}$ and ${\rm offset} - L_{\rm pk}$, one is quite weak (Pearson  coefficient being  $0.07 ^{+ 0.07}_{-0.04}$ with p-value 0.19  for ${\rm offset}-L_{\rm \gamma, iso}$) while the other is relatively quite strong (Pearson  coefficient being  $-0.62 ^{+ 0.03}_{-0.06}$ with p-value 0.02  for ${\rm offset}-L_{\rm pk}$).

Interestingly, the $E_{\rm \gamma,iso}  -  T_{\rm R45,i}$ relation has also been observed, however the coefficients are quite different for the combined GRBs, SGRBs and LGRBs, as shown in Table  \ref{tab:coef} and in Eqs. (\ref{eq:alltr45eiso}) and (\ref{eq:longtr45eiso}). Especially, for the combine sample and for Short GRBs alone, the Pearson coefficients are $0.425 ^{+0.003}_{-0.007}$ with p-value $3.6 \times 10^{-14}$ and $0.53 ^{+0.01}_{-0.04}$ with p-value $0.01$. These relations are quite strong. However, the underlying reason is not clear. 

%\textbf{There is no ${\rm offset} - L_{\rm \gamma,iso}$ correlation for the combined GRBs, SGRBs and LGRBs. But we can find ${\rm offset} - L_{\rm pk}$ correlation for the combined GRBs and SGRBs. It may be ...}

\section{Conclusion and Discussion} \label{sec:discussion}
In this work, we studied the relations of the spatial position of GRBs in their host galaxies with their durations and energies. Using the available data, we tested the correlations between $T_{\rm 90,i}$, $T_{\rm R45,i}$, $E_{\rm \gamma,iso}$, $L_{\rm \gamma, iso}$, $L_{\rm pk}$ and corresponding spatial offset in their host galaxies. We found the host galaxy offsets have negative correlations with the other four parameters. This negative relation for ($T_{\rm 90,i}$ and $T_{\rm R45,i}$) seems reasonable, because LGRBs are most-likely from massive star collapsing, so the GRB positions are  closer to the center of its host galaxies. On the other hand, SGRBs are most-likely from binary mergers. For the binary mergers, the system will experience large velocity kicks at birth, leading to eventual mergers outside the host galaxies \citep{Berger2010}. Therefore, mostly SGRBs have bigger spatial offsets than LGRBs.  But there is no correlation about host galaxy offsets and duration for SGRBs and LGRBs respectively, so different origins offer a natural explanation for the two negative correlations. Interestingly, $L_{\rm pk}$  is related to the offset, while $L_{\rm \gamma, iso}$ is not, and the $E_{\rm \gamma,iso}  -  T_{\rm R45,i}$ relation is quite different of the long GRBs, short GRBs and the combined sample. We also found the special GRBs 170817A and 060218A are outliers in most of the scatter plots, which might indicate there are more subgroups in the GRB samples.

Considering the relation between the $E_{\rm \gamma,iso}$ and the offsets, for the combined GRBs analysis, as the SGRBs have lower energy than LGRBs, it becomes natural that $E_{\rm \gamma,iso}$ and the host galaxy offsets have negative correlation. We have found a similar correlation for SGRBs. We interpret it  as a result of different  progenitors. If the progenitor is NS-BH, the SGRBs have larger $E_{\rm \gamma,iso}$ and smaller spatial offset, because the bigger mass of the system and smaller kick velocity. While for the NS-NS progenitors, the behaviors are on the opposite. For the positive correlation of $E_{\rm \gamma,iso}$ and $T_{\rm R45,i}$, we can find significant correlation for the combined GRBs, SGRBs and LGRBs. Therefore, it is a common property for GRBs, no matter what the central engines or progenitors are. 

We have seen the main tendency is that: for  SGRBs, they are distributed in the outer side of the host galaxies, while for the  LGRBs, they are in the inner side. This supports again that the SGRBs are from mergers of compact stars (NS-NS, or NS-BH), while the LGRBs are from the core collapsing of the massive stars. With this topic, the strongest evidence is the SNe Ic connection to the long GRBs, such as GRB 980425/SN 1998bw \citep{1998Natur.395..670G}, while for the short GRBs is the gravitational wave event association, i.e., the most recently discovered NS-NS merger event GRB 170817A/GW170817 \citep{2017ApJ...848L..12A}. Within the SGRBs, it also shows an anti-correlation between the offset and isotropic $\gamma-$ray energy. This may indicate the double origins for the SGRBs, i.e., NS-NS origin with lower energy and locating nearer the center of the host galaxy, while BH-NS origin with higher energy and locating at outer side. A possible explanation might be that at the birth of the NS or BH, for very high speed NS, the companion compact star has the chance to decouple, while the high speed BH (with more mass) binary can survive, and consequently reaches to outer area of the host galaxy. Also because of the range of the BH masses, it may explain the variety of the magnitude of the kilonovae (or called macronovae, merger novae) \citep{2017ApJ...837...50G}. We have analyzed the correlations by dividing the sample into long GRBs and short GRBs based on the duration being greater or smaller than 2s. It is also possible to perform the same analysis based on the k-means classification, which were given in Section \ref{sec:results}. As one can see, the sample does not change much, and the results may not change much neither.

%There are some shortcomings of this work. 
Our data is biased for strong or optically bright GRBs. The optically dark GRBs are hard to identify the host galaxies and hence the spatial offsets are not available. The GRBs can be dark due to high absorption by the dense medium which may not affect the biasing of offset. In the case intrinsically dark GRBs, the offset might be intrinsically different from the bright GRBs. However to investigate the answer to this property is beyond the scope of this work. It is possible that the GRB location is just occasionally overlapping with a galaxy, while the host galaxy is much dimmer in the foreground or background. This may introduce some bias to the data, especially for the large offset ones. Additionally, our statistical analysis totally rely on observational resolution and technology. 

In the future, one can get more information about the host galaxy with the enhancement of the GRB monitors and observations with higher angular resolution. Monitoring the host galaxies mostly rely on the ground optical follow up efficiency. With more powerful telescope, one can get more information on-site the position of the GRB location (not only the whole galaxy). The environment, such as circum-burst density, star formation rate etc., are also important observations to be made. These information are crucial for understanding the physics of GRBs as well as the final stage of the stellar evolution. 

\section*{Acknowledgments}
This work is supported by the National Basic Research Program of China (973 Program, Grant No. 2014CB845800) and by the National Natural Science Foundation of China (Grant Nos. U1738132, U1231101 and 11773010). RM acknowledges grant by Templeton foundation.

\bibliographystyle{elsarticle-harv} 
\bibliography{msv}

\clearpage

\begin{table}
\centering
\caption{The imputation result for some parameters' error bars. The definitions for RIV, FMI, RE are shown in section(\ref{subsec:imputation}). The subscript 1 and 2 represent positive error bar and negative error bar respectively.}
\begin{tabular}{cccc}
\hline
\multicolumn{1}{c}{Parameter}&
\multicolumn{1}{c}{RIV}&
\multicolumn{1}{c}{FMI}&
\multicolumn{1}{c}{RE}\\
\hline
$T_{\rm 90,i,1}$ & 1.56585E-04 & 1.56573E-04 & 0.99997  \\ 
$T_{\rm 90,i,2}$ & 6.62888E-05 & 6.62866E-05 & 0.99999  \\ 
$T_{\rm R45,i,1}$ & 9.40554E-05 & 9.40510E-05 & 0.99998  \\ 
$T_{\rm R45,i,2}$ & 8.64797E-05 & 8.64759E-05 & 0.99998  \\ 
$E_{\rm \gamma,iso,1}$ & 5.41595E-05 & 5.41581E-05 & 0.99999  \\ 
$E_{\rm \gamma,iso,2}$ & 3.75877E-05 & 3.75870E-05 & 0.99999  \\ 
${\rm offset_ 1}$ & 2.43425E-02 & 2.40396E-02 & 0.99522  \\ 
${\rm offset_ 2}$ & 2.84502E-03 & 2.84096E-03 & 0.99943  \\ 
$L_{\rm \gamma,iso,1}$ & 2.98363E-05 & 2.98358E-05 & 0.99999  \\ 
$L_{\rm \gamma,iso,2}$ & 3.65546E-04 & 3.65479E-04 & 0.99993  \\ 
\hline
\end{tabular}
\label{tab:impu}
\end{table}

\begin{table}
\centering
\caption{Linear coefficients between the four different parameters. All errors are in 1 $\sigma$ confidence level. Pearson, Spearman and Kendall p-value are the mean p-value in the 1 $\sigma$ range of its own coefficient.}
\begin{tabular}{cccccccc}
\hline
\multicolumn{1}{c}{type}&
\multicolumn{1}{c}{values}&
\multicolumn{1}{c}{\tabincell{c} {$\log T_{\rm 90,i}$ \\ vs \\ $\log \rm offset$}}&
\multicolumn{1}{c}{\tabincell{c} {$\log T_{\rm R45,i}$ \\ vs \\ $\log \rm offset$}}&
\multicolumn{1}{c}{\tabincell{c} {$\log E_{\rm \gamma,iso}$ \\ vs \\ $\log \rm offset$}}&
\multicolumn{1}{c}{\tabincell{c} {$\log L_{\rm \gamma,iso}$ \\ vs \\ $\log \rm offset$}}&
\multicolumn{1}{c}{\tabincell{c} {$\log L_{\rm pk}$ \\ vs \\ $\log \rm offset$}}&
\multicolumn{1}{c}{\tabincell{c} {$\log T_{\rm R45,i}$ \\ vs \\ $\log E_{\rm \gamma,iso}$}}\\
\hline
 & Pearson coefficient & $-0.37_{\rm -0.04}^{\rm +0.02}$ & $-0.4_{\rm -0.04}^{\rm +0.02}$ & $-0.29_{\rm -0.04}^{\rm +0.02}$ & $-0.03_{\rm -0.03}^{\rm +0.01}$ & $-0.18_{\rm -0.04}^{\rm +0.02}$ & $0.425_{\rm -0.007}^{\rm +0.003}$ \\  
 & Pearson p-value & 3.9E-05 & 2.8E-05 & 2.1E-03 &  0.73 &  0.07 & 3.6E-14 \\ 
 & Spearman coefficient & $-0.25_{\rm -0.03}^{\rm +0.01}$ & $-0.27_{\rm -0.03}^{\rm +0.01}$ & $-0.2_{\rm -0.03}^{\rm +0.01}$ & $-0.03_{\rm -0.03}^{\rm +0.01}$ & $-0.11_{\rm -0.04}^{\rm +0.01}$ & $0.359_{\rm -0.008}^{\rm +0.003}$ \\ 
\tabincell{c} {All \\ GRBs} & Spearman p-value & 7.3E-03  & 4.1E-03  & 3E-02  & 0.74  & 0.25  & 2.9E-10 \\ 
 & Kendall coefficient & $-0.16_{\rm -0.02}^{\rm +0.01}$ & $-0.18_{\rm -0.02}^{\rm +0.01}$ & $-0.13_{\rm -0.02}^{\rm +0.01}$ & $-0.02_{\rm -0.02}^{\rm +0.01}$ & $-0.08_{\rm -0.03}^{\rm +0.01}$ & $0.249_{\rm -0.005}^{\rm +0.002}$ \\ 
 & Kendall p-value & 1.1E-02 &  6E-03 & 3.7E-02  &  0.72 & 0.22  & 2.6E-10 \\ 
   \hline   
 & Pearson coefficient & $-0.25_{\rm -0.07}^{\rm +0.03}$ & $-0.35_{\rm -0.08}^{\rm +0.03}$ & $-0.36_{\rm -0.08}^{\rm +0.04}$ & $0.07_{\rm -0.04}^{\rm +0.07}$ & $-0.62_{\rm -0.06}^{\rm +0.03}$ & $0.53_{\rm -0.04}^{\rm +0.01}$ \\
\hline
\end{tabular}
\label{tab:coef}
\end{table}

\begin{landscape}
\begin{table}
\caption{Sample for the quantities related to the offset of the GRB host galaxies.  The definition of each parameter is shown in Section \ref{sec:samples}.}
\setlength{\tabcolsep}{0pt}
\centering
\small
\begin{tabular}{ccccccccc}
\hline
 \multicolumn{1}{c}{GRB}&
  \multicolumn{1}{c}{Time}&
  \multicolumn{1}{c}{Instrument}&
 \multicolumn{1}{c}{$z$}&
  \multicolumn{1}{c}{$T_{\rm 90}$}&
  \multicolumn{1}{c}{$T_{\rm R45}$}&
  \multicolumn{1}{c}{$E_{\rm \gamma,iso}$}&
  \multicolumn{1}{c}{offset}&
 \multicolumn{1}{c}{ref}
  \\
  \multicolumn{1}{c}{}&
  \multicolumn{1}{c}{}&
  \multicolumn{1}{c}{}&
 \multicolumn{1}{c}{}&
  \multicolumn{1}{c}{($\rm s$)}&
  \multicolumn{1}{c}{($\rm s$)}&
  \multicolumn{1}{c}{($10^{\rm 52} ~ \rm ergs$)}&
  \multicolumn{1}{c}{($\rm kpc$)}&
 \multicolumn{1}{c}{} \\
 \hline \\
170817A & 12:41:06.47 & Fermi &  0.00968 & $2.0^{\rm +0.5}_{\rm -0.5}$ & ...  &  $4.7E-6^{\rm +7E-7}_{\rm -7E-7}$ & $2.0^{\rm +0.2}_{\rm -0.2}$ & (1);(2);(1);(3)  \\
130907A & 21:41:13 & Swift/KW & 1.238 & $357^{\rm +2}_{\rm -2}$ & $40.31^{\rm +0.14}_{\rm -0.14}$ & $304^{\rm +19}_{\rm -19}$ &  ...  & (4);(5);(5);(4) \\
130831A & 13:04:16 & Swift/KW & 0.4791 & $34.65^{\rm +0.53}_{\rm -0.53}$ & $4.07^{\rm +0.09}_{\rm -0.09}$ & $0.46^{\rm +0.02}_{\rm -0.02}$ & ... & (4);(5);(5);(4)  \\ 
130701A &  04:17:43 & Swift/KW & 1.155 & $4.62^{\rm +0.09}_{\rm -0.09}$ & $1.62^{\rm +0.03}_{\rm -0.03}$ &  $2.60^{\rm +0.09}_{\rm -0.09}$ & ... & (4);(5);(5);(4)  \\ 
130610A & 03:12:13 & Swift/Fermi & 2.092 & $48.5^{\rm +2.4}_{\rm -2.4}$ & $9.3^{\rm +0.3}_{\rm -0.3}$ &  $6.99^{\rm +0.46}_{\rm -0.46}$ &  ... & (4);(5);(5);(4) \\
\hline
\end{tabular}
\\
\begin{flushleft} 
$\rm{^{a}}$ - The errors are imputed by MICE algorithm.\\
$\rm{^{b}}$ - The errors in the original papers are in 90$\%$ confidence level, we changed the errors to 1 $\sigma$ confidence level by multiplying 0.995/1.645. \\
$\rm{^{c}}$ - We changed the values into logarithm form or from the logarithm form into the normal form.
\end{flushleft}
\label{tab:sample}
\end{table}
\end{landscape}

\begin{landscape}
\begin{table}
\caption{Continued from previous page}
\setlength{\tabcolsep}{0pt}
\centering
\small
\begin{tabular}{ccccccccc}
\hline
 \multicolumn{1}{c}{GRB}&
  \multicolumn{1}{c}{Time}&
  \multicolumn{1}{c}{Instrument}&
 \multicolumn{1}{c}{$z$}&
  \multicolumn{1}{c}{$T_{\rm 90}$}&
  \multicolumn{1}{c}{$T_{\rm R45}$}&
  \multicolumn{1}{c}{$E_{\rm \gamma,iso}$}&
  \multicolumn{1}{c}{offset}&
 \multicolumn{1}{c}{ref}
  \\
  \multicolumn{1}{c}{}&
  \multicolumn{1}{c}{}&
  \multicolumn{1}{c}{}&
 \multicolumn{1}{c}{}&
  \multicolumn{1}{c}{($\rm s$)}&
  \multicolumn{1}{c}{($\rm s$)}&
  \multicolumn{1}{c}{($10^{\rm 52} ~ \rm ergs$)}&
  \multicolumn{1}{c}{($\rm kpc$)}&
 \multicolumn{1}{c}{} \\
 \hline \\
 120521C & 23:22:07 & Swift & 6 & $33.2^{\rm +2.4}_{\rm -2.4}$ & $5.32^{\rm +0.24}_{\rm -0.24}$ & $19^{\rm +8}_{\rm -8}$ & ... & (15);(5);(5);(16)  \\ 
120404A & 12:51:02 & Swift & 2.876 & $40.5^{\rm +1.5}_{\rm -1.5}$ & $12.15^{\rm +0.46}_{\rm -0.46}$ &  $4.18^{\rm +0.34}_{\rm -0.34}$ & ... & (4);(5);(5);(4)  \\ 
120327A & 02:55:16 & Swift & 2.813 & $71.2^{\rm +2.3}_{\rm -2.3}$ & $10.20^{\rm +0.33}_{\rm -0.33}$ &  $14.42^{\rm +0.46}_{\rm -0.46}$ & ... & (4);(5);(5);(4)   \\ 
120326A & 01:20:29 & Swift/Fermi & 1.798 & $72.7^{\rm +3.1}_{\rm -3.1}$ & $4.68^{\rm +0.11}_{\rm -0.11}$ &  $3.27^{\rm +0.33}_{\rm -0.33}$ & ... & (4);(5);(5);(4) \\ 
120119A & 04:04:30.21 & Swift/Fermi & 1.728 & $70.4^{\rm +4.3}_{\rm -4.3}$ & $13.6^{\rm +0.2}_{\rm -0.2}$ &  $27.2^{\rm +3.6}_{\rm -3.6}$ & ... & (4);(5);(5);(4)   \\
\hline
\end{tabular}
\\
%\label{tab:sample}
\end{table}
\end{landscape}

\begin{landscape}
\begin{table}
\caption{Continued from previous page}
\setlength{\tabcolsep}{0pt}
\centering
\small
\begin{tabular}{ccccccccc}
\hline
 \multicolumn{1}{c}{GRB}&
  \multicolumn{1}{c}{Time}&
  \multicolumn{1}{c}{Instrument}&
 \multicolumn{1}{c}{$z$}&
  \multicolumn{1}{c}{$T_{\rm 90}$}&
  \multicolumn{1}{c}{$T_{\rm R45}$}&
  \multicolumn{1}{c}{$E_{\rm \gamma,iso}$}&
  \multicolumn{1}{c}{offset}&
 \multicolumn{1}{c}{ref}
  \\
  \multicolumn{1}{c}{}&
  \multicolumn{1}{c}{}&
  \multicolumn{1}{c}{}&
 \multicolumn{1}{c}{}&
  \multicolumn{1}{c}{($\rm s$)}&
  \multicolumn{1}{c}{($\rm s$)}&
  \multicolumn{1}{c}{($10^{\rm 52} ~ \rm ergs$)}&
  \multicolumn{1}{c}{($\rm kpc$)}&
 \multicolumn{1}{c}{} \\
 \hline \\
 091029A & 03:53:22 & Swift & 2.752 & $40.0^{\rm +1.3}_{\rm -1.3}$ & $11.47^{\rm +0.23}_{\rm -0.23}$ &  $7.97^{\rm +0.82}_{\rm -0.82}$ & ... & (4);(5);(5);(4)  \\ 
091024A & 08:56:01 & Swift/Fermi & 1.092 & $115^{\rm +5}_{\rm -5}$ & $24.64^{\rm +0.87}_{\rm -0.87}$ &  $18.4^{\rm +2.0}_{\rm -2.0}$ & ... & (4);(5);(5);(4)   \\ 
091020A & 21:36:44 & Swift/Fermi & 1.71 & $39.0^{\rm +1.1}_{\rm -1.1}$ & $6.00^{\rm +0.17}_{\rm -0.17}$ &  $8.4^{\rm +1.1}_{\rm -1.1}$ & ... & (4);(5);(5);(4)  \\ 
091018A & 20:48:19 & Swift/Konus-RF & 0.97 & $4.44^{\rm +0.15}_{\rm -0.15}$ & $1.240^{\rm +0.022}_{\rm -0.022}$ &  $0.7^{\rm +0.3}_{\rm -0.1}$ & ... & (15);(5);(5);(24)  \\ 
090927 & 10:07:16 & Swift/Fermi & 1.37 & $18.4^{\rm +1.3}_{\rm -1.3}$ & $0.84^{\rm +0.09}_{\rm -0.09}$ &  $0.276^{\rm +0.035}_{\rm -0.035}$ & ... & (4);(5);(5);(4)  \\
\hline
\end{tabular}
\\
%\label{tab:sample}
\end{table}
\end{landscape}

\begin{landscape}
\begin{table}
\caption{Continued from previous page}
\setlength{\tabcolsep}{0pt}
\centering
\small
\begin{tabular}{ccccccccc}
\hline
 \multicolumn{1}{c}{GRB}&
  \multicolumn{1}{c}{Time}&
  \multicolumn{1}{c}{Instrument}&
 \multicolumn{1}{c}{$z$}&
  \multicolumn{1}{c}{$T_{\rm 90}$}&
  \multicolumn{1}{c}{$T_{\rm R45}$}&
  \multicolumn{1}{c}{$E_{\rm \gamma,iso}$}&
  \multicolumn{1}{c}{offset}&
 \multicolumn{1}{c}{ref}
  \\
  \multicolumn{1}{c}{}&
  \multicolumn{1}{c}{}&
  \multicolumn{1}{c}{}&
 \multicolumn{1}{c}{}&
  \multicolumn{1}{c}{($\rm s$)}&
  \multicolumn{1}{c}{($\rm s$)}&
  \multicolumn{1}{c}{($10^{\rm 52} ~ \rm ergs$)}&
  \multicolumn{1}{c}{($\rm kpc$)}&
 \multicolumn{1}{c}{} \\
 \hline \\
 081109A & 07:02:06.61 & Swift/Fermi & 0.9787 & $70.0^{\rm +1.8}_{\rm -1.8}$$\rm{^{b}}$ & $17.20^{\rm +0.42}_{\rm -0.42}$$\rm{^{b}}$ & $1.81^{\rm +0.12}_{\rm -0.12}$ & ... & (4);(25);(25);(4)  \\ 
081029 & 01:43:56.78 & Swift & 3.8479 & $170.0^{\rm +8.5}_{\rm -8.5}$$\rm{^{b}}$ & $49.0^{\rm +2.4}_{\rm -2.4}$$\rm{^{b}}$ & $12.1^{\rm +1.4}_{\rm -1.4}$ & ... & (4);(25);(25);(4) \\ 
081028A & 00:25:00.79 & Swift & 3.038 & $280.0^{\rm +9.1}_{\rm -9.1}$$\rm{^{b}}$ & $78.0^{\rm +1.8}_{\rm -1.8}$$\rm{^{b}}$ & $18.3^{\rm +1.8}_{\rm -1.8}$ & ... &  (4);(25);(25);(4) \\ 
081008A & 19:58:09.38 & Swift/Fermi & 1.969 & $200^{\rm +11}_{\rm -11}$$\rm{^{b}}$ & $32.00^{\rm +0.61}_{\rm -0.61}$$\rm{^{b}}$ & $10^{\rm +1}_{\rm -1}$ & $14.3^{\rm +0.2}_{\rm -0.2}$ &  (4);(25);(25);(4);(6) \\ 
081007A & 05:23:52.74 & Swift/Fermi & 0.5295 & $5.60^{\rm +0.24}_{\rm -0.24}$$\rm{^{b}}$ & $1.600^{\rm +0.061}_{\rm -0.061}$$\rm{^{b}}$ & $0.17^{\rm +0.02}_{\rm -0.02}$ & $0.71^{\rm +0.18}_{\rm -0.18}$ & (4);(25);(25);(4);(6)  \\
\hline
\end{tabular}
\\
%\label{tab:sample}
\end{table}
\end{landscape}

\begin{landscape}
\begin{table}
\caption{Continued from previous page}
\setlength{\tabcolsep}{0pt}
\centering
\small
\begin{tabular}{ccccccccc}
\hline
 \multicolumn{1}{c}{GRB}&
  \multicolumn{1}{c}{Time}&
  \multicolumn{1}{c}{Instrument}&
 \multicolumn{1}{c}{$z$}&
  \multicolumn{1}{c}{$T_{\rm 90}$}&
  \multicolumn{1}{c}{$T_{\rm R45}$}&
  \multicolumn{1}{c}{$E_{\rm \gamma,iso}$}&
  \multicolumn{1}{c}{offset}&
 \multicolumn{1}{c}{ref}
  \\
  \multicolumn{1}{c}{}&
  \multicolumn{1}{c}{}&
  \multicolumn{1}{c}{}&
 \multicolumn{1}{c}{}&
  \multicolumn{1}{c}{($\rm s$)}&
  \multicolumn{1}{c}{($\rm s$)}&
  \multicolumn{1}{c}{($10^{\rm 52} ~ \rm ergs$)}&
  \multicolumn{1}{c}{($\rm kpc$)}&
 \multicolumn{1}{c}{} \\
 \hline \\
 080210A & 07:50:05.43 & Swift & 2.6419 & $44.0^{\rm +4.2}_{\rm -4.2}$$\rm{^{b}}$ & $8.80^{\rm +0.36}_{\rm -0.36}$$\rm{^{b}}$ & $4.77^{\rm +0.29}_{\rm -0.29}$ & ... & (4);(25);(25);(4)   \\ 
080207A & 21:30:21.44 & Swift & 2.0858 & $310.0^{\rm +9.1}_{\rm -9.1}$$\rm{^{b}}$ & $70.0^{\rm +6.1}_{\rm -6.1}$$\rm{^{b}}$ & $16.4^{\rm +1.8}_{\rm -1.8}$ & $6.6^{\rm +3.7}_{\rm -3.7}$ & (4);(25);(25);(4);(6)  \\ 
080129 & 06:06:45.46 & Swift & 4.349 & $46.0^{\rm +2.4}_{\rm -2.4}$$\rm{^{b}}$ & $10.00^{\rm +0.61}_{\rm -0.61}$$\rm{^{b}}$ & $8.0^{\rm +4.2}_{\rm -2.4}$$\rm{^{b}}$ & ... & (25);(25);(25);(25)  \\ 
080123 & 04:21:57.40 & Swift/INTEGRAL & 0.495 & $26.0^{\rm +2.4}_{\rm -2.4}$$\rm{^{b}}$ & $3.10^{\rm +0.48}_{\rm -0.48}$$\rm{^{b}}$ & $0.117^{\rm +0.039}_{\rm -0.039}$ & ... & (4);(25);(25);(4)  \\ 
071227A & 20:13:47.04 & Swift/INTEGRAL/KW & 0.383 & $2.20^{\rm +0.18}_{\rm -0.18}$$\rm{^{b}}$ & $0.500^{\rm +0.036}_{\rm -0.036}$$\rm{^{b}}$ & $0.08^{\rm +0.01}_{\rm -0.01}$ & $15.7^{\rm +0.3}_{\rm -0.3}$ & (7);(25);(25);(4);(6)   \\
\hline
\end{tabular}
\\
%\label{tab:sample}
\end{table}
\end{landscape}

\begin{landscape}
\begin{table}
\caption{Continued from previous page}
\setlength{\tabcolsep}{0pt}
\centering
\small
\begin{tabular}{ccccccccc}
\hline
 \multicolumn{1}{c}{GRB}&
  \multicolumn{1}{c}{Time}&
  \multicolumn{1}{c}{Instrument}&
 \multicolumn{1}{c}{$z$}&
  \multicolumn{1}{c}{$T_{\rm 90}$}&
  \multicolumn{1}{c}{$T_{\rm R45}$}&
  \multicolumn{1}{c}{$E_{\rm \gamma,iso}$}&
  \multicolumn{1}{c}{offset}&
 \multicolumn{1}{c}{ref}
  \\
  \multicolumn{1}{c}{}&
  \multicolumn{1}{c}{}&
  \multicolumn{1}{c}{}&
 \multicolumn{1}{c}{}&
  \multicolumn{1}{c}{($\rm s$)}&
  \multicolumn{1}{c}{($\rm s$)}&
  \multicolumn{1}{c}{($10^{\rm 52} ~ \rm ergs$)}&
  \multicolumn{1}{c}{($\rm kpc$)}&
 \multicolumn{1}{c}{} \\
 \hline \\
 070318A & 07:28:56.08 & Swift & 0.84 & $51.0^{\rm +1.8}_{\rm -1.8}$$\rm{^{b}}$ & $10.3^{\rm +0.3}_{\rm -0.3}$$\rm{^{b}}$ & $3.64^{\rm +0.17}_{\rm -0.17}$ & $0.84^{\rm +0.47}_{\rm -0.47}$ & (4);(32);(32);(4);(6) \\ 
070306A & 16:41:28.17 & Swift & 1.4959 & $260^{\rm +12}_{\rm -12}$$\rm{^{b}}$ & $16.00^{\rm +0.61}_{\rm -0.61}$$\rm{^{b}}$ & $8.26^{\rm +0.41}_{\rm -0.41}$ & $0.78^{\rm +0.39}_{\rm -0.39}$ & (4);(32);(32);(4);(6)  \\ 
070208 & 09:10:34.28 & Swift & 1.165 & $52.00^{\rm +0.61}_{\rm -0.61}$$\rm{^{b}}$ & $4.60^{\rm +0.48}_{\rm -0.48}$$\rm{^{b}}$ & $0.280^{\rm +0.13}_{\rm -0.048}$$\rm{^{b}}$ & $0.83^{\rm +0.34}_{\rm -0.34}$ & (32);(32);(32);(32);(6)  \\ 
070129A & 23:35:10.26 & Swift & 2.34 & $92.0^{\rm +1.8}_{\rm -1.8}$$\rm{^{b}}$ & $28.00^{\rm +0.61}_{\rm -0.61}$$\rm{^{b}}$ & $2.89^{\rm +1.2}_{\rm -0.55}$ & ... & (15);(32);(32);(27) \\ 
070110A & 07:22:41.57 & Swift & 2.3521 & $48.0^{\rm +1.2}_{\rm -1.2}$$\rm{^{b}}$ & $16.20^{\rm +0.54}_{\rm -0.54}$$\rm{^{b}}$ & $5.0^{\rm +0.3}_{\rm -0.3}$ & ... & (4);(32);(32);(4)  \\
\hline
\end{tabular}
\\
%\label{tab:sample}
\end{table}
\end{landscape}

\begin{landscape}
\begin{table}
\caption{Continued from previous page}
\setlength{\tabcolsep}{0pt}
\centering
\small
\begin{tabular}{ccccccccc}
\hline
 \multicolumn{1}{c}{GRB}&
  \multicolumn{1}{c}{Time}&
  \multicolumn{1}{c}{Instrument}&
 \multicolumn{1}{c}{$z$}&
  \multicolumn{1}{c}{$T_{\rm 90}$}&
  \multicolumn{1}{c}{$T_{\rm R45}$}&
  \multicolumn{1}{c}{$E_{\rm \gamma,iso}$}&
  \multicolumn{1}{c}{offset}&
 \multicolumn{1}{c}{ref}
  \\
  \multicolumn{1}{c}{}&
  \multicolumn{1}{c}{}&
  \multicolumn{1}{c}{}&
 \multicolumn{1}{c}{}&
  \multicolumn{1}{c}{($\rm s$)}&
  \multicolumn{1}{c}{($\rm s$)}&
  \multicolumn{1}{c}{($10^{\rm 52} ~ \rm ergs$)}&
  \multicolumn{1}{c}{($\rm kpc$)}&
 \multicolumn{1}{c}{} \\
 \hline \\
 060707A & 21:30:19.49 & Swift & 3.424 & $75.0^{\rm +1.8}_{\rm -1.8}$$\rm{^{b}}$ & $14.00^{\rm +0.61}_{\rm -0.61}$$\rm{^{b}}$ & $4.3^{\rm +1.1}_{\rm -1.1}$ & ... & (4);(32);(32);(4)  \\
060614A & 12:43:48.50 & Swift/KW & 0.125 & $109.00^{\rm +0.61}_{\rm -0.61}$$\rm{^{b}}$ & $27.20^{\rm +0.36}_{\rm -0.36}$$\rm{^{b}}$ & $0.217^{\rm +0.087}_{\rm -0.087}$ & $0.80^{\rm +0.03}_{\rm -0.03}$ & (4);(32);(32);(4);(6)  \\ 
060607A & 05:12:13.35 & Swift & 3.075 & $103.0^{\rm +2.4}_{\rm -2.4}$$\rm{^{b}}$ & $15.10^{\rm +0.48}_{\rm -0.48}$$\rm{^{b}}$ & $11.9^{\rm +2.8}_{\rm -2.8}$ & ... & (4);(32);(32);(4)  \\ 
060605A & 18:15:44.61 & Swift & 3.773 & $19.00^{\rm +0.61}_{\rm -0.61}$$\rm{^{b}}$ & $5.4^{\rm +0.3}_{\rm -0.3}$$\rm{^{b}}$ & $4.23^{\rm +0.61}_{\rm -0.61}$ & $1.35^{\rm +0.86}_{\rm -0.86}$ & (4);(32);(32);(4);(6) \\ 
060604A & 18:19:00.01 & Swift & 2.14 & $40.00^{\rm +0.61}_{\rm -0.61}$$\rm{^{b}}$ & $2.00^{\rm +0.24}_{\rm -0.24}$$\rm{^{b}}$ & $0.50^{\rm +0.12}_{\rm -0.1}$ & ... & (15);(32);(32);(35) \\
\hline
\end{tabular}
\\
%\label{tab:sample}
\end{table}
\end{landscape}

\begin{landscape}
\begin{table}
\caption{Continued from previous page}
\setlength{\tabcolsep}{0pt}
\centering
\small
\begin{tabular}{ccccccccc}
\hline
 \multicolumn{1}{c}{GRB}&
  \multicolumn{1}{c}{Time}&
  \multicolumn{1}{c}{Instrument}&
 \multicolumn{1}{c}{$z$}&
  \multicolumn{1}{c}{$T_{\rm 90}$}&
  \multicolumn{1}{c}{$T_{\rm R45}$}&
  \multicolumn{1}{c}{$E_{\rm \gamma,iso}$}&
  \multicolumn{1}{c}{offset}&
 \multicolumn{1}{c}{ref}
  \\
  \multicolumn{1}{c}{}&
  \multicolumn{1}{c}{}&
  \multicolumn{1}{c}{}&
 \multicolumn{1}{c}{}&
  \multicolumn{1}{c}{($\rm s$)}&
  \multicolumn{1}{c}{($\rm s$)}&
  \multicolumn{1}{c}{($10^{\rm 52} ~ \rm ergs$)}&
  \multicolumn{1}{c}{($\rm kpc$)}&
 \multicolumn{1}{c}{} \\
 \hline \\
 051109A & 01:12:20.00 & Swift/KW & 2.346 & $4.9^{\rm +0.3}_{\rm -0.3}$$\rm{^{b}}$ & $1.300^{\rm +0.061}_{\rm -0.061}$$\rm{^{b}}$ & $6.85^{\rm +0.73}_{\rm -0.73}$ & ... & (4);(32);(32);(4)   \\
051022A & 13:07:58.49 & HETE-2/KW & 0.8 & $178^{\rm +8}_{\rm -8}$ & $10.30^{\rm +0.23}_{\rm -0.23}$ & $56.0^{\rm +5.6}_{\rm -5.6}$ & $1.31^{\rm +0.95}_{\rm -0.95}$ & (4);(39);(42);(4);(6)  \\ 
051016B & 18:28:09.00 & Swift & 0.9364 & $1.400_{\rm -0.061}^{\rm +0.061}$$\rm{^{b}}$ & $0.48^{\rm +0.03}_{\rm -0.03}$$\rm{^{b}}$ & $0.0370^{\rm +0.034}_{\rm -0.0036}$$\rm{^{b}}$ & $2.1^{\rm +1.6}_{\rm -1.6}$ & (32);(32);(32);(32);(6) \\ 
051008A & 16:33:21.00 & Swift & 2.9 & $45.00^{\rm +0.61}_{\rm -0.61}$$\rm{^{b}}$ & $8.60^{\rm +0.24}_{\rm -0.24}$$\rm{^{b}}$ & $9.6^{\rm +1.5}_{\rm -1}$ & ... & (43);(32);(32);(27) \\ 
051001A & 11:11:36.00 & Swift & 2.43 & $56.0_{\rm -1.2}^{\rm +1.2}$$\rm{^{b}}$ & $20.00^{\rm +0.61}_{\rm -0.61}$$\rm{^{b}}$ & $2.09^{\rm +0.56}_{\rm -0.29}$ & ... & (15);(32);(32);(27)  \\
\hline
\end{tabular}
\\
%\label{tab:sample}
\end{table}
\end{landscape}

\begin{landscape}
\begin{table}
\caption{Continued from previous page}
\setlength{\tabcolsep}{0pt}
\centering
\small
\begin{tabular}{ccccccccc}
\hline
 \multicolumn{1}{c}{GRB}&
  \multicolumn{1}{c}{Time}&
  \multicolumn{1}{c}{Instrument}&
 \multicolumn{1}{c}{$z$}&
  \multicolumn{1}{c}{$T_{\rm 90}$}&
  \multicolumn{1}{c}{$T_{\rm R45}$}&
  \multicolumn{1}{c}{$E_{\rm \gamma,iso}$}&
  \multicolumn{1}{c}{offset}&
 \multicolumn{1}{c}{ref}
  \\
  \multicolumn{1}{c}{}&
  \multicolumn{1}{c}{}&
  \multicolumn{1}{c}{}&
 \multicolumn{1}{c}{}&
  \multicolumn{1}{c}{($\rm s$)}&
  \multicolumn{1}{c}{($\rm s$)}&
  \multicolumn{1}{c}{($10^{\rm 52} ~ \rm ergs$)}&
  \multicolumn{1}{c}{($\rm kpc$)}&
 \multicolumn{1}{c}{} \\
 \hline \\
 050315 & 20:59:42.00 & Swift & 1.95 & $95.0^{\rm +1.8}_{\rm -1.8}$$\rm{^{b}}$ & $19.30^{\rm +0.48}_{\rm -0.48}$$\rm{^{b}}$ & $6.15^{\rm +0.30}_{\rm -0.30}$ & $0.98^{\rm +0.48}_{\rm -0.48}$ & (4);(32);(32);(4);(6)  \\ 
050223A & 03:09:06.00 & Swift/INTEGRAL & 0.5915 & $17.40^{\rm +0.54}_{\rm -0.54}$$\rm{^{b}}$ & $5.1^{\rm +0.3}_{\rm -0.3}$$\rm{^{b}}$ & $0.0700^{\rm +0.03}_{\rm -0.0061}$$\rm{^{b}}$ & ... & (32);(32);(32);(32)  \\ 
050126 & 12:00:53.00 & Swift & 1.29 & $29.0^{\rm +1.2}_{\rm -1.2}$$\rm{^{b}}$ & $8.10^{\rm +0.48}_{\rm -0.48}$$\rm{^{b}}$ & $2.47^{\rm +0.25}_{\rm -0.25}$ & ... & (4);(32);(32);(4)  \\ 
041006A & 12:18:08 & HETE-2 & 0.716 & 18$\rm{^{a}}$ & $4.26^{\rm +0.12}_{\rm -0.12}$ & $3.11^{\rm +0.89}_{\rm -0.89}$ & $2.57^{\rm +0.06}_{\rm -0.06}$ & (4);(11);(42);(4);(6)  \\ 
040924A & 11:52:11 & HETE-2/KW & 0.859 & 2.39$\rm{^{a}}$ & $0.49^{\rm +0.02}_{\rm -0.02}$ & $0.98^{\rm +0.10}_{\rm -0.10}$ & $2.23^{\rm +0.12}_{\rm -0.12}$ & (4);(11);(42);(4);(6)  \\
\hline
\end{tabular}
\\
%\label{tab:sample}
\end{table}
\end{landscape}

\begin{table}
\caption{Continued from previous page}
\setlength{\tabcolsep}{0pt}
\centering
\small
\begin{tabular}{ccccccccc}
\hline
 \multicolumn{1}{c}{GRB}&
  \multicolumn{1}{c}{Time}&
  \multicolumn{1}{c}{Instrument}&
 \multicolumn{1}{c}{$z$}&
  \multicolumn{1}{c}{$T_{\rm 90}$}&
  \multicolumn{1}{c}{$T_{\rm R45}$}&
  \multicolumn{1}{c}{$E_{\rm \gamma,iso}$}&
  \multicolumn{1}{c}{offset}&
 \multicolumn{1}{c}{ref}
  \\
  \multicolumn{1}{c}{}&
  \multicolumn{1}{c}{}&
  \multicolumn{1}{c}{}&
 \multicolumn{1}{c}{}&
  \multicolumn{1}{c}{($\rm s$)}&
  \multicolumn{1}{c}{($\rm s$)}&
  \multicolumn{1}{c}{($10^{\rm 52} ~ \rm ergs$)}&
  \multicolumn{1}{c}{($\rm kpc$)}&
 \multicolumn{1}{c}{} \\
 \hline \\
971214B & 23:20:41.2 & BeppoSAX/CGRO & 3.42 & $30.0^{\rm +6.3}_{\rm -6.3}$ & $1.45^{\rm +0.39}_{\rm -0.31}$ &  $22.1^{\rm +2.7}_{\rm -2.7}$ & $1.11^{\rm +0.56}_{\rm -0.56}$ & (4);(48);(50);(4);(52)  \\ 
970828A & 17:44:36.8 & RXTE/CGRO/Konus & 0.958 & 147$\rm{^{a}}$ & 12.9$\rm{^{a}}$ & $30.4^{\rm +3.6}_{\rm -3.6}$ & ... & (4);(36);(49);(4)  \\ 
970508A & 21:41:46.8 & BeppoSAX/CGRO & 0.835 & $14.0^{\rm +3.6}_{\rm -3.6}$ & $1.70^{\rm +0.31}_{\rm -0.29}$ &  $0.65^{\rm +0.13}_{\rm -0.13}$ & $0.091^{\rm +0.090}_{\rm -0.090}$ & (4);(48);(50);(4);(52)  \\ 
970228A & 02:58:01 & BeppoSAX/Konus & 0.695 & $56.0^{\rm +3.2}_{\rm -3.2}$ & $1.54^{\rm +0.18}_{\rm -0.19}$ &  $1.65^{\rm +0.16}_{\rm -0.16}$ & $3.1^{\rm +0.2}_{\rm -0.2}$ & (4);(48);(50);(4);(6) \\ 
     \hline
\end{tabular}
\\
\label{tab:samplelast}
\end{table}

References. 
(1) \citet{Pozanenko2017};
(2) \citet{Abbott2017};
(3) \citet{AbbottA2017};
(4) \citet{Ruffini2016};
(5) \citet{Golkhou2014};
(6) \citet{Li2016ApJS};
(7) \citet{Sang2016};
(8) \citet{Jeong2014};
(9) \citet{Dichiara2016};
(10) \citet{Deng2016};
(11) \citet{Cano2017};
(12) \citet{Liang2015};
(13) \citet{Turpin2016};
(14) \citet{Siellez2016};
(15) \citet{Racusin2016};
(16) \citet{Song2016};
(17) \citet{Von2014};
(18) \citet{Margutti2012};
(19) \citet{Tunnicliffe2014};
(20) \citet{Kopac2012};
(21) \citet{Wang2016};
(22) \citet{Virgili2012};
(23) \citet{Lyman2017};
(24) \citet{Sakamoto2011};
(25) \citet{Butler2010};
(26) \citet{Arcodia2016};
(27) \citet{Wang2014};
(28) \citet{Robertson2012};
(29) \citet{Perley2016A};
(30) \citet{Qin2013B};
(31) \citet{Wei2014};
(32) \citet{Butler2007};
(33) \citet{Dado2016};
(34) \citet{Cui2012};
(35) \citet{Lv2014A};
(36) \citet{Chandra2012};
(37) \citet{KruhlerAA2015};
(38) \citet{Troja2008};
(39) \citet{Pelangeon2008};
(40) \citet{Kopac2013};
(41) \citet{Rykoff2009};
(42) \citet{Collazzi2008};
(43) \citet{Perley2013};
(44) \citet{Yonetoku2010};
(45) \citet{Minaev2014};
(46) \citet{Melandri2008};
(47) \citet{Sakamoto2005};
(48) \citet{Frontera2009};
(49) \citet{Guidorzi2005};
(50) \citet{Rossi2008};
(51) \citet{Barniol2014};
(52) \citet{Bloom2002};
(53) \url{https://gammaray.nsstc.nasa.gov/batse/grb/catalog/current/index.html}

\begin{table}
\caption{$L_{\rm pk}$ and the related spectral information. The definition of each parameter is shown in Section \ref{sec:samples}. The data are selected with accompanied offset available.}
\setlength{\tabcolsep}{0pt}
\centering
\small
\begin{tabular}{ccccccc}
\hline
 \multicolumn{1}{c}{GRB}&
  \multicolumn{1}{c}{$L_{\rm pk}$}&
%  \multicolumn{1}{c}{offset}&
  \multicolumn{1}{c}{$-\alpha$}&
  \multicolumn{1}{c}{$-\beta$}&
  \multicolumn{1}{c}{$E_{\rm pk}$}&
  \multicolumn{1}{c}{Model}&
 \multicolumn{1}{c}{ref}
  \\
  \multicolumn{1}{c}{}&
  \multicolumn{1}{c}{($\rm 10^{\rm 52} ~ erg ~ s^{\rm -1}$)}&
 % \multicolumn{1}{c}{($\rm kpc$)}&
   \multicolumn{1}{c}{}&
  \multicolumn{1}{c}{}&
  \multicolumn{1}{c}{($\rm keV$)}&
   \multicolumn{1}{c}{}&
 \multicolumn{1}{c}{} \\
 \hline \\
170817A & $1.6E-5^{\rm +2.5E-5}_{\rm -4E-6}$$\rm{^{f}}$ %& $2.0^{\rm +0.2}_{\rm -0.2}$ 
& ... & ... & ... &  & (1);(2)  \\
130606A & $8^{\rm +2.8}_{\rm -2.1}$$\rm{^{c,d}}$ %& $0.36^{\rm +0.12}_{\rm -0.12}$ 
& $1.140^{\rm +0.091}_{\rm -0.091}$$\rm{^{b}}$ & ... & $294^{\rm +54}_{\rm -30}$$\rm{^{b}}$ & CPL & (3);(4);(4);(4)  \\ 
130427A & $7.1^{\rm +0.3}_{\rm -0.3}$$\rm{^{c,d}}$ %& $4.1^{\rm +0.1}_{\rm -0.1}$ 
& $0.91^{\rm +0.01}_{\rm -0.01}$ & $3.18^{\rm +0.03}_{\rm -0.03}$ & $877.8^{\rm +4.9}_{\rm -4.9}$ & Band & (3);(4);(5);(5);(5)  \\ 
110731A & $8.1^{\rm +1.1}_{\rm -1.1}$$\rm{^{c,d}}$ %& $1.47^{\rm +0.08}_{\rm -0.08}$ 
& $0.82^{\rm +0.03}_{\rm -0.03}$ & $2.32^{\rm +0.03}_{\rm -0.02}$ & $312^{\rm +29}_{\rm -27}$$\rm{^{e}}$ & Band & (3);(4);(6);(6);(6)  \\ 
100816A & $0.984^{\rm +0.022}_{\rm -0.022}$ %& $8.2^{\rm +1.4}_{\rm -1.4}$$\rm{^{b}}$ 
& ... & ... & ... &  & (7);(8)  \\ 
091208B & $2.05^{\rm +0.31}_{\rm -0.41}$ %& 0.79$\rm{^{a}}$ 
& ... & ... & ... &  & (9);(10)   \\ 
091127A & $0.358^{\rm +0.015}_{\rm -0.015}$ %& $1.35^{\rm +0.30}_{\rm -0.30}$ 
& ... & ... & ... &  & (9);(4)  \\ 
090618 & $0.937^{\rm +0.012}_{\rm -0.012}$ %& $4.5^{\rm +0.3}_{\rm -0.3}$ 
& ... & ... & ... &  & (7);(4)   \\ 
090426 & $2.90^{\rm +0.78}_{\rm -0.63}$$\rm{^{c,d}}$ %& $0.49^{\rm +0.25}_{\rm -0.25}$ 
& $1.93^{\rm +0.13}_{\rm -0.13}$$\rm{^{b}}$ & ... & ... & SPL & (3);(4);(4) \\ 
090424A & $1.680^{\rm +0.029}_{\rm -0.029}$ %& $2.62^{\rm +0.24}_{\rm -0.24}$ 
& ... & ... & ... &  & (7);(4)  \\ 
090418A & $1.18^{\rm +0.19}_{\rm -0.34}$ % & 0.74$\rm{^{a}}$ 
& ... & ... & ... &  & (9);(10) \\ 
090205A & $1.30^{\rm +1.4}_{\rm -0.94}$$\rm{^{c,d}}$ %& $2.7^{\rm +2.0}_{\rm -2.0}$ 
& $1.0^{\rm +0.6}_{\rm -1.2}$$\rm{^{b}}$ & ... & $33^{\rm +19}_{\rm -19}$$\rm{^{b}}$ & CPL & (3);(4);(11);(11)  \\ 
090102A & $5.83^{\rm +0.84}_{\rm -0.82}$ %& $0.78^{\rm +0.52}_{\rm -0.52}$ 
& ... & ... & ... &  & (9);(4) \\ 
081221A & $11.8^{\rm +0.2}_{\rm -0.2}$ %& $3.3^{\rm +0.6}_{\rm -0.6}$ 
& ... & ... & ... &  & (7);(4) \\ 
081121A & $5.37^{\rm +0.74}_{\rm -0.74}$ %& $2^{\rm +1}_{\rm -1}$ 
& ... & ... & ... &  & (7);(4) \\ 
081008A & $0.55^{\rm +0.14}_{\rm -0.14}$$\rm{^{c,d}}$ %& $14.3^{\rm +0.2}_{\rm -0.2}$ 
& $1.010^{\rm +0.073}_{\rm -0.073}$$\rm{^{b}}$ & $2.09^{\rm +0.13}_{\rm -0.13}$$\rm{^{b}}$ & $166^{\rm +22}_{\rm -22}$$\rm{^{b}}$ & Band & (3);(4);(12);(12);(12) \\ 
081007A & $0.034^{\rm +0.016}_{\rm -0.014}$$\rm{^{c,d}}$ %& $0.71^{\rm +0.18}_{\rm -0.18}$ 
& $1.40^{\rm +0.36}_{\rm -0.6}$$\rm{^{b}}$ & ... & $27.0^{\rm +6.7}_{\rm -9.7}$$\rm{^{b}}$ & CPL & (3);(4);(11);(11)  \\ 
080928 & $0.474^{\rm +0.18}_{\rm -0.084}$$\rm{^{c,d}}$ %& $14.8^{\rm +0.3}_{\rm -0.3}$ 
& $1.70^{\rm +0.06}_{\rm -0.06}$$\rm{^{b}}$ & ... & $74^{\rm +147}_{\rm -16}$$\rm{^{b}}$ & CPL & (3);(4);(11);(11) \\ 
080916A & $0.108^{\rm +0.005}_{\rm -0.005}$ %& 0.14$\rm{^{a}}$ 
& ... & ... & ... &  & (7);(10) \\ 
080805A & $0.211^{\rm +0.11}_{\rm -0.051}$$\rm{^{c,d}}$ %& $4.0^{\rm +0.4}_{\rm -0.4}$ 
& $1.540^{\rm +0.054}_{\rm -0.054}$$\rm{^{b}}$ & ... & $300^{\rm +333}_{\rm -115}$$\rm{^{b}}$ & CPL & (3);(4);(11);(11) \\ 
080707A & $0.221^{\rm +0.076}_{\rm -0.073}$$\rm{^{c,d}}$ %& $0.76^{\rm +0.46}_{\rm -0.46}$ 
& $1.78^{\rm +0.17}_{\rm -0.17}$ & ... & ... & SPL & (3);(4);(13)  \\
080607A & $129^{\rm +26}_{\rm -26}$ %& $5.4^{\rm +1.9}_{\rm -1.9}$ 
& ... & ... & ... &  & (9);(4)  \\ 
080605A & $9.82^{\rm +0.18}_{\rm -0.18}$ %& $0.70^{\rm +0.27}_{\rm -0.27}$ 
& ... & ... & ... &  & (7);(4)  \\ 
080430 & $0.103^{\rm +0.13}_{\rm -0.007}$ %& $0.81^{\rm +0.37}_{\rm -0.37}$ 
& ... & ... & ... &  & (14);(4)  \\ 
080319C & $6.04^{\rm +8}_{\rm -0.42}$ %& $7.4^{\rm +0.4}_{\rm -0.4}$ 
& ... & ... & ... &  & (14);(4)  \\   
 080319B & $7.39^{\rm +0.71}_{\rm -0.71}$ & ... & ... & ... &  & (9);(4)  \\ 
080207A & $0.61^{\rm +0.11}_{\rm -0.11}$ & ... & ... & ... &  & (7);(4)  \\ 
071227A & $0.0334^{\rm +0.0049}_{\rm -0.0049}$ & ... & ... & ... &  & (15);(4)   \\ 
071122 & $0.035^{\rm +0.06}_{\rm -0.026}$$\rm{^{c,d}}$  & $1.60^{\rm +0.24}_{\rm -0.24}$$\rm{^{b}}$ & ... & $111^{\rm +318}_{\rm -41}$$\rm{^{b}}$ & CPL & (3);(4);(11);(11)   \\ 
 071112C & $4.0^{\rm +1.1}_{\rm -1.1}$$\rm{^{c,d}}$  & $1.090^{\rm +0.042}_{\rm -0.042}$$\rm{^{b}}$ & ... & ... & SPL & (3);(4);(16) \\
071010A & $1.97^{\rm +0.9}_{\rm -0.78}$$\rm{^{c,d}}$  & $2.24^{\rm +0.19}_{\rm -0.22}$$\rm{^{b}}$  & ... & ... & SPL & (3);(10);(4) \\
      \hline
\end{tabular}
\\
\begin{flushleft} 
$\rm{^{a}}$ - The errors are imputed by MICE algorithm.\\
$\rm{^{b}}$ - The errors in the original papers are in 90$\%$ confidence level, we changed the errors to 1 $\sigma$ confidence level by multiplying 0.995/1.645. \\
$\rm{^{c}}$ - The values are calculated using the spectral values in order to change the energy band. \\
$\rm{^{d}}$ - We changed the values into logarithm or from logarithm into the normal form. \\
$\rm{^{e}}$ - The values are calculated using $\alpha$ and break energy $E_{0}$. \\
$\rm{^{f}}$ - The peak luminosity of GRB 170817A is derived from 50 $\rm ms$ time resolution light curve. \\
\end{flushleft}
\label{tab:lpk}
\end{table}

\begin{table}
\caption{Continued from previous page}
\setlength{\tabcolsep}{0pt}
\centering
\small
\begin{tabular}{ccccccc}
\hline
 \multicolumn{1}{c}{GRB}&
  \multicolumn{1}{c}{$L_{\rm pk}$}&
  \multicolumn{1}{c}{$-\alpha$}&
  \multicolumn{1}{c}{$-\beta$}&
  \multicolumn{1}{c}{$E_{\rm pk}$}&
  \multicolumn{1}{c}{Model}&
 \multicolumn{1}{c}{ref}
  \\
  \multicolumn{1}{c}{}&
  \multicolumn{1}{c}{($\rm 10^{\rm 52} ~ erg ~ s^{\rm -1}$)}&
  %\multicolumn{1}{c}{($\rm kpc$)}&
   \multicolumn{1}{c}{}&
  \multicolumn{1}{c}{}&
  \multicolumn{1}{c}{($\rm keV$)}&
   \multicolumn{1}{c}{}&
 \multicolumn{1}{c}{} \\
 \hline \\
071010B & $0.605^{\rm +0.014}_{\rm -0.014}$ & ... & ... & ... &  & (7);(4) \\
070802A & $0.24^{\rm +0.2}_{\rm -0.14}$$\rm{^{c,d}}$  & $1.80^{\rm +0.18}_{\rm -0.18}$$\rm{^{b}}$ & ... & $55^{\rm +161}_{\rm -21}$$\rm{^{b}}$ & CPL & (3);(4);(11);(11)  \\
 070724A & $0.0158^{\rm +0.0034}_{\rm -0.0014}$  & ... & ... & ... &  & (15);(4)  \\ 
070714B & $1.45^{\rm +0.17}_{\rm -0.38}$ & ... & ... & ... &  & (9);(4)  \\ 
070508 & $2.374^{\rm +0.037}_{\rm -0.037}$  & ... & ... & ... &  & (7);(4)  \\ 
070429B & $0.246^{\rm +0.038}_{\rm -0.038}$  & ... & ... & ... &  & (15);(17)  \\
070318A & $0.073^{\rm +0.048}_{\rm -0.015}$$\rm{^{c,d}}$ & $1.410^{\rm +0.048}_{\rm -0.048}$$\rm{^{b}}$ & ... & $196^{\rm +269}_{\rm -47}$$\rm{^{b}}$ & CPL & (3);(4);(18);(18) \\ 
070306A & $0.867^{\rm +1.4}_{\rm -0.027}$  & ... & ... & ... &  & (14);(4)  \\ 
061217 & $0.108^{\rm +0.018}_{\rm -0.018}$ & ... & ... & ... &  & (15);(19) \\ 
061210A & $0.215^{\rm +0.014}_{\rm -0.014}$  & ... & ... & ... &  & (15);(19) \\ 
061110B & $0.59^{\rm +0.94}_{\rm -0.38}$$\rm{^{c,d}}$  & $0.70^{\rm +0.12}_{\rm -0.12}$$\rm{^{b}}$ & ... & $550^{\rm +593}_{\rm -151}$$\rm{^{b}}$ & CPL & (3);(4);(18);(18)  \\ 
061110A & $0.0170^{\rm +0.011}_{\rm -0.0051}$$\rm{^{c,d}}$ & $1.60^{\rm +0.06}_{\rm -0.06}$$\rm{^{b}}$ & ... & $106^{\rm +181}_{\rm -24}$$\rm{^{b}}$ & CPL & (3);(4);(18);(18)   \\ 
061007A & $14.3^{\rm +2.3}_{\rm -1.8}$ & ... & ... & ... &  & (9);(10)  \\ 
061006A & $0.2460^{\rm +0.012}_{\rm -0.0077}$  & ... & ... & ... &  & (15);(4)  \\ 
060912A & $1.076^{\rm +0.16}_{\rm -0.095}$$\rm{^{c}}$ & $5.2^{\rm +1.2}_{\rm -1.2}$  & ... & $200^{\rm +339}_{\rm -67}$$\rm{^{b}}$ & CPL & (20);(4);(18);(18) \\ 
060801A & $0.476^{\rm +0.062}_{\rm -0.016}$  & ... & ... & ... &  & (15);(17) \\ 
060729A & $0.0196^{\rm +0.0085}_{\rm -0.0041}$$\rm{^{c,d}}$  & $1.80^{\rm +0.06}_{\rm -0.06}$$\rm{^{b}}$ & ... & $67^{\rm +139}_{\rm -15}$$\rm{^{b}}$ & CPL & (3);(4);(18);(18) \\ 
060719A & $0.54^{\rm +0.16}_{\rm -0.15}$$\rm{^{c,d}}$  & $1.90^{\rm +0.06}_{\rm -0.06}$$\rm{^{b}}$ & ... & $55^{\rm +31}_{\rm -31}$$\rm{^{b}}$ & CPL & (3);(4);(18);(18)  \\ 
060614A & $0.00808^{\rm +0.0013}_{\rm -0.00094}$ & ... & ... & ... &  & (9);(4)  \\ 
060605A & $0.91^{\rm +0.13}_{\rm -0.13}$ & ... & ... & ... &  & (21);(4) \\ 
060602A & $0.061^{\rm +0.025}_{\rm -0.008}$  & ... & ... & ... &  & (15);(10)  \\ 
060522 & $2.33^{\rm +0.78}_{\rm -0.81}$ & ... & ... & ... &  & (9);(4)   \\ 
060502B & $0.0065^{\rm +9E-4}_{\rm -9E-4}$ & ... & ... & ... &  & (15);(19)  \\
060418A & $1.96^{\rm +0.43}_{\rm -0.13}$  & ... & ... & ... &  & (14);(4)   \\ 
060218A & $4.3E-6^{\rm +1.5E-6}_{\rm -1.7E-6}$ & ... & ... & ... &  & (9);(4)  \\ 
060206A & $7.18^{\rm +0.27}_{\rm -0.27}$  & ... & ... & ... &  & (7);(4)  \\ 
 060124A & $10.2^{\rm +3.1}_{\rm -2.9}$   & ... & ... & ... &  & (9);(4)   \\ 
060121A & $24.5^{\rm +1.6}_{\rm -1.6}$  & ... & ... & ... &  & (15);(4)   \\ 
060115A & $1.55^{\rm +0.13}_{\rm -0.13}$  & ... & ... & ... &  & (7);(4)   \\ 
051221A & $0.258^{\rm +0.009}_{\rm -0.009}$  & ... & ... & ... &  & (15);(4)  \\ 
051022A & $3.12^{\rm +0.41}_{\rm -0.41}$  & ... & ... & ... &  & (9);(4)  \\ 
051016B & $0.065^{\rm +0.048}_{\rm -0.038}$   & ... & ... & ... &  & (9);(4) \\ 
050904A & $8.5^{\rm +2.3}_{\rm -2.4}$  & ... & ... & ... &  & (9);(4)  \\ 
050826A & $0.0033^{\rm +0.0032}_{\rm -8E-4}$  & ... & ... & ... &  & (15);(4)  \\ 
 050824A & $0.18^{\rm +0.19}_{\rm -0.17}$$\rm{^{c,d}}$   & $2.76^{\rm +0.23}_{\rm -0.23}$$\rm{^{b}}$ & ... & ... & SPL & (3);(4);(4)  \\ 
050820A & $3.22^{\rm +0.3}_{\rm -0.45}$   & ... & ... & ... &  & (9);(4)  \\ 
050813 & $0.04^{\rm +0.02}_{\rm -0.02}$  & ... & ... & ... &  & (15);(17)  \\ 
050724A & $0.0099^{\rm +0.0023}_{\rm -0.001}$  & ... & ... & ... &  & (15);(4)  \\ 
050709A & $0.0540^{\rm +0.0067}_{\rm -0.0069}$  & ... & ... & ... &  & (15);(4)  \\ 
050525A & $1.159^{\rm +0.016}_{\rm -0.016}$  & ... & ... & ... &  & (7);(4)  \\ 
050509B & $7E-4^{\rm +0.001}_{\rm -5E-4}$  & ... & ... & ... &  & (15);(4)  \\ 
050416A & $0.087^{\rm +0.019}_{\rm -0.015}$  & ... & ... & ... &  & (9);(4)  \\ 
 050401A & $12.0^{\rm +2.3}_{\rm -2.6}$   & ... & ... & ... &  & (9);(4)  \\ 
     \hline
\end{tabular}
\\
%\label{tab:lpk1}
\end{table}

\begin{table}
\caption{Continued from previous page}
\setlength{\tabcolsep}{0pt}
\centering
\small
\begin{tabular}{ccccccc}
\hline
 \multicolumn{1}{c}{GRB}&
  \multicolumn{1}{c}{$L_{\rm pk}$}&
  \multicolumn{1}{c}{$-\alpha$}&
  \multicolumn{1}{c}{$-\beta$}&
  \multicolumn{1}{c}{$E_{\rm pk}$}&
  \multicolumn{1}{c}{Model}&
 \multicolumn{1}{c}{ref}
  \\
  \multicolumn{1}{c}{}&
  \multicolumn{1}{c}{($\rm 10^{\rm 52} ~ erg ~ s^{\rm -1}$)}&
  %\multicolumn{1}{c}{($\rm kpc$)}&
   \multicolumn{1}{c}{}&
  \multicolumn{1}{c}{}&
  \multicolumn{1}{c}{($\rm keV$)}&
   \multicolumn{1}{c}{}&
 \multicolumn{1}{c}{} \\
 \hline \\
050315 & $0.88^{\rm +0.33}_{\rm -0.31}$   & ... & ... & ... &  & (9);(4)  \\ 
041006A & $0.531^{\rm +0.022}_{\rm -0.022}$   & ... & ... & ... &  & (9);(4)  \\ 
040924A & $0.86^{\rm +0.09}_{\rm -0.09}$  & ... & ... & ... &  & (9);(4)  \\ 
030429A & $0.75^{\rm +0.19}_{\rm -0.22}$   & ... & ... & ... &  & (9);(4)  \\ 
030323A & $1.14^{\rm +0.61}_{\rm -0.51}$  & ... & ... & ... &  & (9);(4)  \\ 
021004A & $0.91^{\rm +0.19}_{\rm -0.19}$   & ... & ... & ... &  & (9);(4)  \\ 
010921A & $0.13^{\rm +0.01}_{\rm -0.01}$   & ... & ... & ... &  & (21);(4)   \\ 
010222A & $28.7^{\rm +1.3}_{\rm -1.2}$  & ... & ... & ... &  & (9);(4)  \\ 
000210A & $7.0^{\rm +2.8}_{\rm -1.8}$   & ... & ... & ... &  & (9);(4) \\ 
991208A & $4.68^{\rm +1.2}_{\rm -0.83}$   & ... & ... & ... &  & (9);(4)  \\ 
990705A & $2.21^{\rm +0.19}_{\rm -0.17}$ & ... & ... & ... &  & (9);(4)   \\ 
990510A & $5.2^{\rm +1.1}_{\rm -1.4}$   & ... & ... & ... &  & (9);(4)  \\
990123A & $27.5^{\rm +1.2}_{\rm -1.2}$   & ... & ... & ... &  & (9);(4)   \\
980703A & $1.20^{\rm +0.13}_{\rm -0.23}$  & ... & ... & ... &  & (9);(4)   \\ 
980613A & $0.193^{\rm +0.068}_{\rm -0.057}$   & ... & ... & ... &  & (9);(4)  \\ 
971214B & $9.5^{\rm +5.3}_{\rm -4}$   & ... & ... & ... &  & (9);(22)  \\ 
970508A & $0.164^{\rm +0.021}_{\rm -0.021}$  & ... & ... & ... &  & (9);(22)  \\ 
970228A & $1.54^{\rm +0.21}_{\rm -0.12}$   & ... & ... & ... &  & (9);(4) \\ 
     \hline
\end{tabular}
\label{tab:lpklast}

References.                                  
(1) \citet{Zhang5851};                
(2) \citet{AbbottA2017};        
(3) \citet{Deng2016};           
(4) \citet{Li2016ApJS};                 
(5) \citet{Goldstein2016};                
(6) \citet{Ackermann2013};                      
(7) \citet{Lin2016};
(8) \citet{Kopac2012};
(9) \citet{Yonetoku2010};
(10) \citet{Lyman2017};
(11) \citet{Butler2010};
(12) \citet{Heussaff2013};
(13) \citet{Krimm2009};
(14) \citet{Ukwatta2010};
(15) \citet{Zhang2009};
(16) \citet{Nava2012};
(17) \citet{Tunnicliffe2014};
(18) \citet{Butler2007};
(19) \citet{Cui2012};
(20) \citet{Rizzuto2007};
(21) \citet{Rossi2008};
(22) \citet{Bloom2002}
\end{table}

\end{document}